# Accounting statement analysis at industry level. A gentle introduction to the compositional approach

Germà Coenders[1]
Department of Economics.
University of Girona,
Girona, Spain

Núria Arimany Serrat
Department of Economics and Business.
University of Vic - Central University of Catalonia,
Vic, Spain.

**Abstract**

Compositional data are contemporarily defined as positive vectors, the ratios among whose elements are of interest to the researcher. Financial statement analysis by means of accounting ratios a.k.a. financial ratios fulfils this definition to the letter. Compositional data analysis solves the major problems in statistical analysis of standard financial ratios at industry level, such as skewness, non-normality, non-linearity, outliers, and dependence of the results on the choice of which accounting figure goes to the numerator and to the denominator of the ratio. Despite this, compositional applications to financial statement analysis are still rare. In this article, we present some transformations within compositional data analysis that are particularly useful for financial statement analysis. We show how to compute industry or sub-industry means of standard financial ratios from a compositional perspective by means of geometric means. We show how to visualise firms in an industry with a compositional principal-component-analysis biplot; how to classify them into homogeneous financial performance profiles with compositional cluster analysis; and how to introduce financial ratios as variables in a statistical model, for instance to relate financial performance and firm characteristics with compositional regression models. We show an application to the accounting statements of Spanish wineries using the decomposition of return on equity by means of DuPont analysis, and a step-by-step tutorial to the compositional freeware CoDaPack.

[1] Corresponding author. Facultat de Ciències Econòmiques i Empresarials. C. Universitat de Girona 10. 17003 Girona, Spain. germa.coenders@udg.edu

## 1. Financial ratios as carriers of relative information

The relative nature of *financial ratios*, a.k.a. *accounting ratios* (i.e., ratios comparing selected figures in financial statements), enables them to evaluate the firm's position compared to its counterparts in the industry or to itself over time, taking into account differences or evolution in firm size (e.g., Ross et al., 2003). Financial ratios have been used in practical management performance and strategic assessment and in research relating them to other financial or non-financial variables (Altman, 1968; Amat Salas, 2020; Barnes, 1987; Faello, 2015; Horrigan, 1968; Nugroho and Dewayanto, 2025; Qin et al., 2022; Soukal et al., 2024; Staňková and Hampel, 2023; Tascón et al., 2018; Veganzones and Severin, 2021; Willer do Prado et al., 2016) including, for instance, fraud detection, stock market returns, and firm survival, default or bankruptcy. A typical example of financial ratio is that of assets over equity. This ratio tells how indebted a firm is, since assets can be decomposed into equity and liabilities, i.e., debt. It is also referred to as the leverage ratio.

While the use of standard financial ratios in diagnostics of the financial health of single firms (or comparing e.g. two firms) is straightforward, analysing a sample of firms from an industry requires statistical methods. Using standard financial ratios as variables in statistical analyses within an industry has been related to a number of serious problems, such as *asymmetry* (Faello, 2015; Frecka and Hopwood, 1983; Linares-Mustarós, et al., 2018; Oktaviano et al., 2024; Trejo-Pech et al., 2023), connected to it severe *non-normality* of the distributions (Adcock et al., 2015; Buijink and Jegers, 1986; Deakin, 1976; Durana et al., 2025; Iotti et al., 2023; 2024a; 2024b; 2024c; Lueg et al., 2014; Martikainen et al., 1995; McLeay and Omar, 2000; So, 1987; Valaskova et al., 2023), *heteroskedasticity* (Barnes, 1986; Faello, 2015; McDonald and Morris, 1984; Mulet-Forteza et al., in press), *non-linearity* of the relationships (Balcaen and Ooghe, 2006; Carreras-Simó and Coenders, 2021; Cowen and Hoffer, 1982; Keasey and Watson, 1991; Taoushianis, 2025), extreme *outliers* (Deshpande, 2023; Ezzamel and Mar-Molinero, 1990; Frecka and Hopwood, 1983; Kane et al., 1998; Lev and Sunder, 1979; Liu et al., 2025; Magrini, 2025; McLeay, 1986; Nyitrai and Virág, 2019; Oktaviano et al., 2024; Watson, 1990), and even dependence of the results on the arbitrary decision regarding which accounting figure appears in the numerator and which in the denominator of the ratio (Coenders et al., 2023a; Frecka and Hopwood, 1983; Linares-Mustarós, et al., 2022). The results of many statistical analyses are invalid when all or some of these problems occur and the results and conclusions of said analyses can be affected to a great extent. It has long been known that these problems not only have serious consequences for single ratios but also when combining ratios into composite indicators by means of factor analysis and related methods (Cowen and Hoffer, 1982; Martikainen et al., 1995). Readers unfamiliar with the statistical consequences of asymmetry, outliers, non-normality, heteroskedasticity and non-linearity can resort to any introductory statistics or econometrics handbook.

These serious problems have also been reported in other scientific fields using ratios (Isles, 2020). It must be noted that when financial ratios were first developed in the late 19th century (Brown Sister, 1955; Horrigan, 1968), statistical analysis methods were in their infancy, which speaks by itself for the fact that financial ratios were not conceived with a statistical usage in mind. The theory on *ratio measurement scales* (Stevens, 1946) had not even been developed. This situation has changed dramatically, with a large body of statistical and econometrical research in the field of accounting being available nowadays (Gruszczyński, 2022). As a matter of fact, the first use of the term *econometrics* was made by Paweł Ciompa in 1910 in the accounting field (Ciompa, 1910).

Up to now, only rarely has the accounting research community recognised the importance of these problems, and when this has been the case, diverse ad-hoc solutions have been applied on the spot for a problem at a time. For instance, outliers have been dealt with by just removing them (Ezzamel and Mar-Molinero, 1990; Frecka and Hopwood, 1983; Kubenka and Myskova, 2022; Magrini, 2025; Martikainen et al., 1995; Martin-Melero et al., 2025; So, 1987; Watson, 1990) or replacing them with the nearest non-outlying values (Demiraj et al., 2024; Deshpande, 2023; Duong et al., 2026; Geng et al., 2026; Gupta, 2024; Lev and Sunder, 1979; Laus et al., 2025; Liu et al., 2025; Mlawu et al., 2026; Naz et al., 2023; Nyitrai and Virág, 2019; Vu et al., 2023). The number of identified outliers in at least one ratio has in some cases neared one third of all firms, and these procedures have thus seriously impaired the sample representativeness. Asymmetry has been dealt with by using transformations such as the square root, the cubic root or generalised polynomials (Akbar et al., 2025; Deakin, 1976; Ezzamel and Mar-Molinero, 1990; Frecka and Hopwood, 1983; Martikainen et al., 1995; Taoushianis, 2025), their generalization as Box-Cox transformations (Mcleay and Omar, 2000; Watson, 1990), or by ignoring the original ratio values and considering only their rank order (Cheng and Fang, 2025; Kane et al., 1998; Lueg et al., 2014). Non-normality has been dealt with non-parametric statistics (Durana et al., 2025; Hazami-Ammar, 2024; Iotti et al., 2023; 2024a; 2024b; 2024c; Latief and Suhendah, 2023; Valaskova et al., 2023) or more complex statistical models (Adcock et al., 2015; Trejo-Pech et al., 2023). Heteroskedasticity has been dealt with weighted least squares methods (Musallam, 2018) or robust standard errors (Geng et al., 2026). Rather than that, here we present a simple unified approach to deal with all problems simultaneously which makes it possible to use the full sample without removing outliers and is compatible with any statistical method from the simplest to the most complex, and not just with a limited range of them as the case is for non-parametric methods.

Financial ratios constitute a genuine case of researchers' and professionals' interest in relative rather than absolute accounting figures and thus a natural field of application of *Compositional Data* (CoDa) analysis, which has the same objective. Essentially, CoDa are arrays of strictly positive numbers for which ratios between them are considered to be relevant (Egozcue and Pawlowsky-Glahn, 2019) which perfectly fits the notion of financial statement analysis. The CoDa methodology offers a number of advantages in statistical analysis of financial statements, as compared to standard financial ratios. Among other features, CoDa treat accounting figures in a symmetric fashion in such a way that results do not depend on numerator and denominator permutation. CoDa also tend to reduce outliers, heteroskedasticity, and non-normality, and to linearize relationships. Far from being a statistical refinement, the CoDa methodology leads to very substantial differences in the analysis results whenever it has been compared with standard financial ratios (Arimany-Serrat et al., 2022; Carreras-Simó and Coenders, 2021; Coenders et al., 2023a; Coenders and Arimany-Serrat, 2025a, 2025b; Creixans-Tenas et al., 2019; Dao et al., 2026; Escaramís and Arbussà, in press; Hernandez Romero and Coenders, 2025; Jofre-Campuzano and Coenders, 2022; Keivani et al., 2026; Linares-Mustarós et al., 2018; 2022; Magrini, 2025; Mulet-Forteza et al., in press).

Since the seminal works by Aitchison (1982, 1986), CoDa analysis has become a well-established methodology, notably present in accessible textbooks (van den Boogaart and Tolosana-Delgado, 2013; Filzmoser et al., 2018; Greenacre, 2018; Pawlowsky-Glahn et al., 2015) and software (Calle et al., 2023; Comas-Cufí and Thió-Henestrosa, 2011; van den Boogaart and Tolosana-Delgado, 2013; Filzmoser et al., 2018; Greenacre, 2018; Palarea-Albaladejo and Martín-Fernández, 2015; Thió-Henestrosa and Martín-Fernández, 2005), and continues to be further developed well after forty years (Coenders et al., 2023b; Greenacre et al., 2023).

The study of the relative importance of chemical elements in geological analysis spurred most of the early interest in CoDa (Aitchison, 1986; Buccianti et al., 2006). Nowadays, CoDa are being used in the social sciences and economics in general (Coenders and Ferrer-Rosell, 2020; Fry, 2011; Martínez-Garcia et al., 2023) and in finance in particular, to answer research questions concerning the relative importance of magnitudes. Financial examples include crowdfunding (Davis et al., 2017), bond ratings (Tallapally, 2009), financial markets (Camacho et al., 2026; Kokoszka et al., 2019; Li et al., 2019; Ortells et al., 2016; Vega-Baquero and Santolino, 2022a; Wang et al., 2019), municipal budgeting (Voltes-Dorta et al., 2014), insurance (Belles-Sampera et al., 2016; Boonen et al., 2019; Fiori and Rosazza Gianin, 2025; Gan and Valdez, 2021; Verbelen et al., 2018), exchange rates (Gámez-Velázquez and Coenders, 2020; Maldonado et al., 2021a; 2021b), banking (Vega-Baquero and Santolino, 2022b), portfolios (Glassman and Riddick, 1996; Joueid and Coenders, 2018; Vega-Gámez and Alonso-González, 2024), systemic risk (Fiori and Coenders, 2025; Fiori and Porro, 2023; Porro, 2022), household finance (Fry et al., 1996; 2000; 2001; Gokhale et al., 2024; Mclaren et al., 1995; Tian et al., 2024), intermediary market share (Dyhrberg et al., 2025), and equity ownership structure (Ahmed et al., 2023). The application to accounting and financial statement analysis spans about a decade (Arimany-Serrat and Coenders, 2025; Arimany-Serrat et al., 2022; 2023; Arimany-Serrat and Sgorla, 2024; Carreras-Simó and Coenders, 2020; 2021; Coenders, 2025, Coenders et al., 2023a; Coenders and Arimany-Serrat, 2025a; 2025b; Creixans-Tenas et al., 2019; Dao et al., 2026; Escaramís and Arbussà, in press; Farreras-Noguer et al., 2025; Hernandez Romero and Coenders, 2025; Jofre-Campuzano and Coenders, 2022; Keivani et al., 2026; Linares-Mustarós et al., 2018; 2022; Magrini, 2025; Molas-Colomer et al., 2024; 2026; Mulet-Forteza et al., in press; Rondós-Casas et al., 2026; Saus-Sala et al., 2021; 2023; 2024).

This article starts explaining why and how financial statements and financial ratios should be considered as CoDa, including the necessary transformations. *DuPont analysis*, a very simple case of financial statement analysis is used as storyline. Then, the dataset of an example in the Spanish winery industry is presented, with microdata in Appendix 1. Next, four approaches to compositional industry analysis are illustrated with the example data. The first three approaches deal with the financial statements in themselves. Financial statements are summarised, visualised, and classified. The fourth approach is devoted to establishing relationships between financial indicators, non-financial indicators and other firm or management characteristics. This is accomplished by introducing the transformed financial ratios as variables in statistical models. The final section concludes. A software guide is included in Appendix 2 and alternative approaches are presented in Appendices 3 and 5. A comparison with standard financial ratios is presented in appendix 4.

## 2. Financial statements as compositional data

A $D$−part composition is defined as an array of $D$ strictly positive numbers, called *parts*, the relative magnitude of which is of interest to the researcher (Aitchison, 1986):

$$\mathbf{x} = \left(x_1, x_2, ..., x_D\right) \text{ with } x_j > 0,\ j = 1, 2, ..., D, \qquad (1)$$

Some rules have to be followed in order to introduce accounting figures in financial statements in a *D*–part composition, which boil down to avoiding negative accounting figures and their overlap (Creixans-Tenas et al., 2019):

- Even if sometimes financial ratios involve accounting figures which may be negative, its use is advised against in the financial literature, because they can cause a discontinuity, outliers, or even a reversal of interpretation when the accounting figure which may be negative is in the denominator (Lev and Sunder, 1979; Linares-Mustarós et al., 2022). Negative accounting figures are also advised against from the point of view of *measurement theory*. Computing a ratio is a meaningful operation only for variables in a *ratio scale*, which need to have a meaningful absolute zero (Stevens, 1946) and thus no negative values.
  In general, accounting figures are negative because they imply the subtraction of other positive accounting figures, which are the ones to be used. This means, for instance, that when building ratios, one should directly use revenues and costs rather than profit or current assets and current liabilities rather than working capital. This limitation implies no loss of information whatsoever. For instance, a ratio conveying the same information as the standard *margin* ratio (profit/revenues) can be constructed from only the non-negative revenue and cost figures. Let $x_1$=revenues, $x_2$=costs, $x_3$=$x_1$–$x_2$=profit. The always positive revenues over costs ratio ($x_1/x_2$) can easily be shown to be just a transformation of the problematic profit over revenues ratio ($x_3/x_1$):

$$\frac{x_1}{x_2} = \frac{x_1}{x_1 - x_3} = \frac{1}{\frac{x_1 - x_3}{x_1}} = \frac{1}{1 - \frac{x_3}{x_1}}\ . \qquad (2)$$

- It must also be taken into account that parts may not overlap. For instance, one could not use $x_4$: assets and $x_5$: non-current assets because $x_5$ is a part of $x_4$. In compositional data terminology, $x_4$: assets are an *amalgamation* of $x_5$: non-current assets and $x_6$: current assets. Using both amalgamations and their constituent parts is extremely problematic (Pawlowsky-Glahn et al., 2015). Rather, the choice between using only the amalgamation or only the individual parts should be made at the problem definition stage and cannot be changed afterwards (van den Boogaart and Tolosana-Delgado, 2013). It is not essential to use all constituent parts, which is referred to as a *subcomposition* in compositional data terminology. Accordingly, the feasible choices to handle $x_4$ to $x_6$ are: a) to use only $x_4$; b) to use $x_5$ and $x_6$; c) to use only $x_5$; and d) to use only $x_6$.

The ultimate choice of parts will depend on the analysis objectives or research questions. The researchers will in principle like to select the accounting figures needed to compute their favourite financial ratios and refine the choice by avoiding overlap and subtraction. In the example we use in this article, the parts represented by the $x_j$ variables are the following *D*=4 positive and non-overlapping financial statement account categories:

- $x_1$: revenues,
- $x_2$: costs,
- $x_3$: liabilities,
- $x_4$: assets.

These account categories are very relevant in practice because they make it possible to compute the common profitability, *turnover*, margin, and *leverage* ratios in classical DuPont analysis. DuPont analysis was developed in 1914 by Donaldson Brown (Dale et al., 1980), and owes its name to the firm where he was working at that time. It has continued to be in use ever since as a popular method for decomposing the firm's performance into a small number of financial ratios (Baležentis et al, 2019; Chen et al., 2014; Saus-Sala et al., 2021; 2023).

According to DuPont analysis, firm profitability can be decomposed as the product of turnover, margin and leverage, according to the following standard financial ratios, some of which are computed from accounting figures which may be negative.

- Turnover measures the efficiency of a firm's use of its assets in generating sales revenue:

$$\text{Turnover}=\text{revenues/assets}=x_1/x_4. \qquad (3)$$

- Margin is the percentage of sales revenue that is turned into profit:

$$\text{Margin}=\text{profit/revenues}=(x_1-x_2)/x_1. \qquad (4)$$

- Leverage measures assets generated per unit of shareholders equity. It is also a measure of indebtedness:

$$\text{Leverage}=\text{assets/equity}=x_4/(x_4-x_3). \qquad (5)$$

- The return on equity (ROE) is a common measure of profitability defined as:

$$\text{ROE}=\text{profit/equity}=(x_1-x_2)/(x_4-x_3), \qquad (6)$$

  and can be decomposed as:

$$\text{ROE}=\text{turnover}\times\text{margin}\times\text{leverage}. \qquad (7)$$

High margin and turnover values are always desirable. An excessive leverage can result in an excessive indebtedness and make the firm vulnerable. In addition, leverage also multiplies margin when margin is negative. For these reasons, leverage is a less preferred manner of attaining high ROE figures.

DuPont analysis is chosen as storyline in this article both because of its simplicity (with only $D$=4 account categories involving a handful of ratios) and its popularity. It goes without saying that account categories could be considered in greater detail, by increasing both the number of accounting figures $D$ and the set of feasible financial ratios computed from them, which would make for a more detailed financial statement analysis (see Appendix 5).

## 3. CoDa transformations

### 3.1 Pairwise log-ratios

The usual approach to statistical analysis of CoDa is to use existent standard statistical methods on transformed data. *Logarithms of ratios* are the standard transformation in CoDa (Aitchison, 1986). The simplest case of a log-ratio is that between only two accounting figures (*pairwise log-ratios*, e.g., Creixans-Tenas et al., 2019; Coenders and Arimany-Serrat, 2025a; 2025b; Farreras-Noguer et al., 2025; Greenacre, 2018; 2019; Keivani et al., 2026; Mulet-Forteza et al., in press; Saus-Sala et al., 2021) and can also be understood as the log difference between the two:

$$\log\left(\frac{x_1}{x_2}\right)=\log\left(x_1\right)-\log\left(x_2\right). \qquad (8)$$

Unlike a standard ratio, which is bounded between zero and infinity, a log-ratio is symmetric in the sense that its range is from minus infinity to plus infinity, the whole *real line*, making it a *real variable*. This has three key advantages.

- It coincides with the support of the normal probability distribution. It is clear that a bounded variable cannot possibly be normally distributed. Although there is never the guarantee that an unbounded real variable will be normally distributed, it is often the case for log-ratios in empirical data sets (Aitchison, 1982). From a theoretical point of view, the normality of a pairwise log-ratio results from the joint log-normality of the two parts involved, and there is a compositional version of the central-limit theorem which states that variables which are the result of many small independent causes acting multiplicatively are log-normal (Pawlowsky-Glahn et al., 2015).
- The linear prediction functions in linear regression (see section 8) are also unbounded. When fitting a standard ratio as dependent variable in a linear regression model, some predicted values could actually be impossible values below zero. The prediction for a log-ratio will never be an impossible value. Some financial ratios are actually fractions of a total and also have an upper bound equal to 1, which compounds the problem.
- Asymmetry is the main reason for heteroskedasticity, as the variance of standard ratios is unavoidable higher at the upper than at the lower end (Faello, 2015).

Besides, a log-ratio is symmetric in the sense that permuting the numerator and denominator parts leads to the same distance from zero and affects no other property of the log-ratio than the sign (Linares-Mustarós et al., 2022):

$$\log\left(\frac{x_1}{x_2}\right)=\log\left(x_1\right)-\log\left(x_2\right)=-\left(\log\left(x_2\right)-\log\left(x_1\right)\right)=-\log\left(\frac{x_2}{x_1}\right). \qquad (9)$$

For instance, the correlation of an external non-financial indicator with a permuted log-ratio equals the correlation with the original log-ratio with a reversed sign. This property does not hold for standard financial ratios. Correlating $x_1/x_2$ with a non-financial indicator can give conflicting results with respect to correlating it with $x_2/x_1$ (Coenders et al., 2023a; Linares-Mustarós et al., 2022). There is no other reason than agreement to use one ratio or its permutation. For a single firm, the fact that $x_1/x_2$=0.5 provides the same information as the fact that $x_2/x_1$=2. However, in statistical analyses at industry level, the results of the one and the other ratio may stand in contradiction.

Finally, if one of the accounting figures being compared in the ratio is close to zero, it may lead to an outlying standard ratio when placed in the denominator and to a typical ratio when placed in the numerator (Ezzamel and Mar-Molinero, 1990; Frecka and Hopwood, 1983; Kane et al., 1998; Lev and Sunder, 1979; Martikainen et al., 1995). For log-ratios, placement makes no difference (Coenders et al., 2023a; Linares-Mustarós et al., 2018; 2022; Molas-Colomer et al., 2024).

Table 1 shows a toy example of seven fictional firms and two accounting figures $x_1$ and $x_2$. For ease of computation, we show logarithms to base 10 represented as $\log_{10}(x)$, which just tell how many times 10 has to be multiplied by itself in order to get the desired value. $\log_{10}(1{,}000{,}000)=6$ because $10^6=1{,}000{,}000$. Since $10^0 = 1$, $\log_{10}(1) = 0$. There is perfect symmetry around 1 and 0: $0.000001 = 1/1{,}000{,}000 = 1/10^6 = 10^{-6}$, so that $\log_{10}(0.000001) = -6$.

The interpretation of the log-ratios with respect to standard ratios is straightforward. When $x_1/x_2$ is larger than 1, $\log_{10}(x_1/x_2)$ is positive. When $x_1/x_2$ is smaller than 1, $\log_{10}(x_1/x_2)$ is negative. The larger $x_1/x_2$, the larger $\log_{10}(x_1/x_2)$.

Note that, like ratios, logarithms focus on relative differences between firms. Ratios and logarithms are thus mutually compatible (Stevens, 1946) and should be routinely used together for data in a ratio scale, when the meaningful difference between two figures is relative, meaning that it lies in their ratio and not in their subtraction. Oddly enough, in the financial statement analysis literature, logs are mostly used for variables measuring firm size in absolute terms, like the number of employees.

For example, if we take firms 3,4, and 5 in the toy example in Table 1 ($x_2$ values 100, 1,000 and 10,000), in relative terms, the difference between 1,000 and 100, which is 1,000/100=10, is the same as the relative difference between 10,000 and 1,000, which is 10,000/1,000=10. Accordingly, their log differences 3−2=1 and 4−3=1 are the same. Once the logarithm has been applied, subtraction is meaningful again. Subtraction is an essential operation in statistics. For instance, the residual is the actual value minus the predicted value, the variance is based on the subtraction of the mean from the actual value, etc.

Note that the values of $x_1$ and $x_2$ in Table 1 are fully symmetrical while the standard ratios $x_1/x_2$ and $x_2/x_1$ are not symmetrical at all. In the ratio $x_1/x_2$, firms 1 and 2 appear as outliers and, in the ratio $x_2/x_1$, firms 6 and 7. In the ratio $x_1/x_2$, firms 4, 5, 6, and 7 are concentrated in the very short [0, 1] interval. In the ratio $x_2/x_1$ the same holds for firms 1, 2, 3, and 4.

Conversely, the logarithms of the ratios $\log_{10}(x_2/x_1)$ and $\log_{10}(x_1/x_2)$ are fully symmetrical, have no outliers, and permutation of numerator and denominator only leads to a sign reversal.

| Firm | $x_1$ | $x_2$ | $x_2/x_1$ | $x_1/x_2$ | $\log_{10}(x_1)$ | $\log_{10}(x_2)$ | $\log_{10}(x_2/x_1)$ | $\log_{10}(x_1/x_2)$ |
|---|---|---|---|---|---|---|---|---|
| 1 | 1,000,000 | 1 | 0.000001 | 1,000,000 | 6 | 0 | –6 | 6 |
| 2 | 100,000 | 10 | 0.0001 | 10,000 | 5 | 1 | –4 | 4 |
| 3 | 10,000 | 100 | 0.01 | 100 | 4 | 2 | –2 | 2 |
| 4 | 1,000 | 1,000 | 1 | 1 | 3 | 3 | 0 | 0 |
| 5 | 100 | 10,000 | 100 | 0.01 | 2 | 4 | 2 | –2 |
| 6 | 10 | 100,000 | 10,000 | 0.0001 | 1 | 5 | 4 | –4 |
| 7 | 1 | 1,000,000 | 1,000,000 | 0.000001 | 0 | 6 | 6 | –6 |

**Table 1.** Toy example with seven firms

Natural logarithms (to base $e$=2.718281828…) represented as log($x$) are commoner in economics and finance, are the ones used in most compositional software, and will be used from here on, but any base could be used without affecting the properties of CoDa analysis.

Some log-ratios between pairs of accounting figures are especially interesting in DuPont analysis. By definition, turnover compares revenues and assets:

$$y_1 = \log\left(\frac{x_1}{x_4}\right). \tag{10}$$

As shown in Equation (2), comparing revenues and costs provides a notion of margin:

$$y_2 = \log\left(\frac{x_1}{x_2}\right). \tag{11}$$

In the same vein, comparing liabilities and assets provides a notion of leverage. Even if it does not correspond with the standard leverage definition in Equation (5), higher liabilities with respect to assets do imply higher leverage:

$$y_3 = \log\left(\frac{x_3}{x_4}\right). \tag{12}$$

Indeed, $x_3/x_4$ is just a transformation of the standard leverage in the form of a pairwise ratio and without a denominator which may be negative for some firms:

$$\frac{x_3}{x_4} = \frac{x_4 - x_4 + x_3}{x_4} = 1 - \frac{x_4 - x_3}{x_4} = 1 - \frac{1}{\dfrac{x_4}{x_4 - x_3}}. \tag{13}$$

It must be noted that ROE involves all four accounting figures and cannot be expressed as a pairwise log-ratio. It cannot either be computed as the product $y_1y_2y_3$.

Potentially, $D(D–1)/2$ different pairwise log-ratios can be computed, although some of them may fail to have any financial interpretation or theoretical interest, ratio choice becoming a potentially problematic issue. Great care must also be taken to prevent ratios from being mutually redundant, meaning that the information of some ratios is already contained in other (Barnes, 1987; Chen and Shimerda, 1981; Magrini, 2025; Pohlman and Hollinger, 1981). For instance, in the above example, a log-ratio computed as the ability of revenues to pay for liabilities $y_4 = \log\left(x_1/x_3\right)$ would be equal to $y_1 - y_3$ :

$$
\begin{aligned}
&y_1 - y_3 = \log\left(\frac{x_1}{x_4}\right) - \log\left(\frac{x_3}{x_4}\right) = \log\left(x_1\right) - \log\left(x_4\right) - \left(\log\left(x_3\right) - \log\left(x_4\right)\right) = \\
&\log\left(x_1\right) - \log\left(x_3\right) = \log\left(\frac{x_1}{x_3}\right) = y_4.
\end{aligned} \tag{14}
$$

Some guidelines to prevent redundancy in pairwise log-ratios are given by Greenacre (2019) and applied by Creixans-Tenas et al. (2019) in the financial-statement context. Greenacre (2019) recommends drawing a graph in which the accounting figures are *vertices* (*nodes*), and the log-ratios are *connections* (*edges*). The graph must necessarily be connected and acyclic. This means that:

- It is possible to join any two accounting figures following the connections (i.e., the log-ratios).
- There may not be closed circuits, that is, when following the edges of the graph from one accounting figure to any other, no accounting figure can be visited twice. In other words, there is only one possible path to join any two accounting figures.

It can be proven by contradiction that such a graph has exactly $D-1$ edges (i.e., log-ratios). If it has fewer edges, it cannot connect all accounting figures, and if it has more edges then there must be a cycle (Greenacre, 2019). $D$–1 pairwise log-ratios so chosen can be proven to contain all information about the $D$-part compositional dataset, in other words all information about the relative importance of the $D$ accounting figures.

While any graph fulfilling these conditions will do the job, statistically speaking, it is good practice to use a graph with substantive interpretation, based on expert knowledge or in the light of the research purpose. In our DuPont analysis case, we want log-ratios to be related to the concepts of turnover, margin and leverage, namely $y_1$, $y_2$ and $y_3$, which fortunately fulfil the conditions according to the connected acyclic graph in the top panel of Fig. 1. Edges can be drawn as arrows without affecting the graph properties, the arrows pointing at the numerator of the log-ratio for clarification purposes only. In other words, accounting figures are considered joined even when going against the arrow directions. See Appendix 5 for further examples.

As an example of inappropriate log-ratio choice, when substituting $y_4 = \log\left(x_1/x_3\right)$ in Equation (14) for $y_2$, there would be a cycle connecting $x_1$, $x_3$ and $x_4$ (in other words, one could go from $x_1$ to $x_3$ either directly or through $x_4$) while $x_2$ would not be connected to any of the other parts (bottom panel in Fig. 1).

Users must be warned that there may be more than one way to choose a sensible set of $D$–1 interpretable and non-redundant pairwise log-ratios, and the results of some statistical analyses which are based on distances (e.g., biplots, principal component analysis and cluster analysis as used in Sections 6 and 7) depend on this choice (Hron et al., 2021). These distance-based statistical methods require alternative log-ratios, as shown in Section 3.2. Conversely, pairwise log-ratios are appropriate as input variables for statistical models (Section 8).

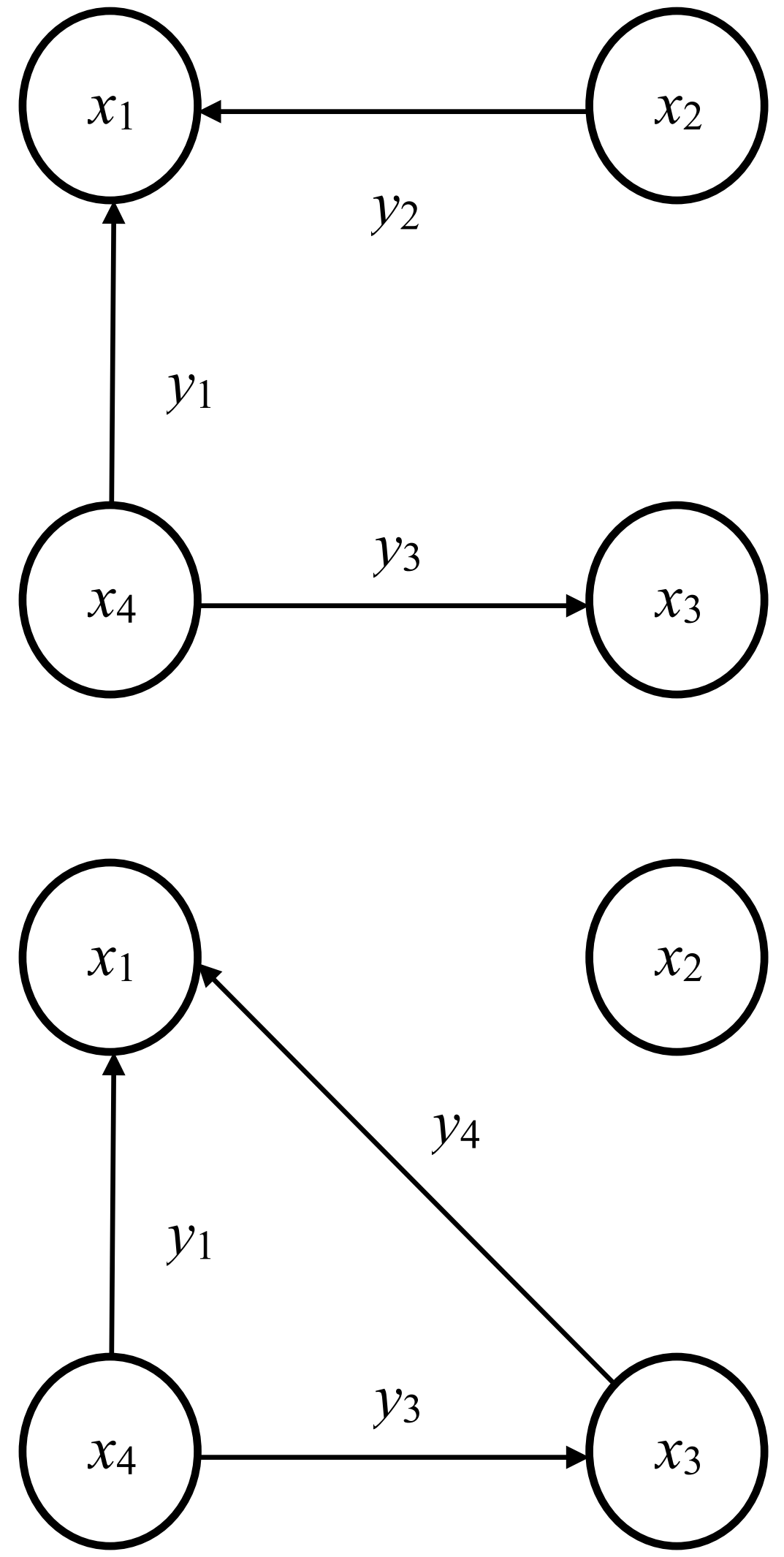

**Figure 1.** Connected acyclic graph (top). Unconnected cyclic graph (bottom)

### 3.2 Centred log-ratios

Pairwise log-ratios are not the only possibility in the CoDa methodology. This methodology can completely dispense with log-ratio choice by ensuring that $D$ so-called *centred log-ratios* or clr (Aitchison, 1983) also contain all information about the relative importance of $D$ accounting figures. Any log-ratio the researcher might be interested in is a function of these $D$ centred log-ratios. Centred log-ratios compare each part, in the numerator, with the *geometric mean* of all parts for each individual firm, in the denominator. They have no accounting interpretation in themselves, but they are used as raw data in multivariate descriptive analysis methods such as cluster analysis, principal component analysis, and biplots, as shown in Sections 6 and 7:

$$clr_j = \log\left(\frac{x_j}{\sqrt[D]{x_1 x_2 \ldots x_D}}\right) \quad \text{with } j = 1, 2, \ldots, D. \tag{15}$$

In our DuPont example we would have four centred log-ratios:

$$
\begin{aligned}
clr_1 &= \log\left(\frac{x_1}{\sqrt[4]{x_1x_2x_3x_4}}\right)\\
clr_2 &= \log\left(\frac{x_2}{\sqrt[4]{x_1x_2x_3x_4}}\right)\\
clr_3 &= \log\left(\frac{x_3}{\sqrt[4]{x_1x_2x_3x_4}}\right)\\
clr_4 &= \log\left(\frac{x_4}{\sqrt[4]{x_1x_2x_3x_4}}\right).
\end{aligned}
\tag{16}
$$

All possible pairwise log-ratios are contained in the centred log-ratios. Note how $y_1$ can be obtained from $clr_1$ and $clr_4$:

$$
\begin{aligned}
&clr_1 - clr_4 = \log\left(\frac{x_1}{\sqrt[4]{x_1x_2x_3x_4}}\right) - \log\left(\frac{x_4}{\sqrt[4]{x_1x_2x_3x_4}}\right) =\\
&\log(x_1) - \log\left(\sqrt[4]{x_1x_2x_3x_4}\right) - \left(\log(x_4) - \log\left(\sqrt[4]{x_1x_2x_3x_4}\right)\right) =\\
&\log(x_1) - \log(x_4) = \log\left(\frac{x_1}{x_4}\right) = y_1.
\end{aligned}
\tag{17}
$$

As we will show in Sections 6 and 7, even if using centred log-ratios as raw data, the interpretation can revert to the more easily interpretable pairwise log-ratios or even to standard financial ratios (Saus-Sala et al., 2021; 2023; 2024).

### 3.3. Zero replacement

A commonly mentioned limitation of CoDa is that the accounting figures of interest may contain no zero values for log-ratios to be computed (Martín-Fernández et al., 2011). However, a fact that is often overlooked is that exactly the same holds for standard financial ratios: a zero accounting figure is not relative to anything and thus the ratio is not a feasible operation according to measurement theory (Stevens, 1946). The ratio is used to measure how many times a magnitude contains another, and this has no answer when one of the magnitudes is zero. If the zero value is the denominator, the standard ratio cannot even be computed.

Unlike the case in standard financial ratio analysis, CoDa include an advanced toolbox for *zero imputation* (a.k.a. *zero replacement*) prior to log-ratio computation under the most common assumptions (Martín Fernández et al., 2012). This provides CoDa with a head advantage compared to standard financial ratio analysis in the presence of zeros and ultimately makes financial statement analysis possible even when some accounting figures of interest equal zero. Shortly stated, zeros are replaced with a meaningful small value following certain statistical properties.

The need for some form of zero treatment was recognised at the very beginning of the development of the CoDa methodology (Aitchison, 1982). See Mariadassou and Coenders, (2025); and Coenders et al. (2023b) for recent reviews on the topic. From early simple methods (Aitchison, 1986; Fry et al., 2000; Martín-Fernández et al., 2003), developments moved to advanced methods, with Palarea-Albaladejo and Martín-Fernández (2008; 2015); and Martín Fernández et al. (2011; 2012) being key references.

In the literature of compositional financial statement analysis, the most popular imputation method by far is the *log-ratio Expectation-Maximization (EM) method* (Palarea-Albaladejo and Martín-Fernández, 2008). This method is similar to the standard EM method for imputing missing data, as the imputed value is predicted from the available values with a statistical model. However, in the compositional case it adds the restriction that the imputed values must be "small". In particular, they are constrained to be below the minimum observed value of each part or below any other value specified by the user, called *detection limit*.

User-defined detection limits are particularly useful in the following case. If the minimum non-zero value corresponds to a firm with a very low figure, replacement below this limit could cause replaced values to be outliers. In this case, we recommend setting the detection limit to be a bit higher. From our experience, detection limits around the mean of the 5 % lowest non-zero values tend to work well.

The zero imputation methodologies require the number of zero values to be small, ideally below 20 % for any of the accounting figures (Palarea-Albaladejo and Martín-Fernández, 2008). Before imputation, percentages of zeros should thus be examined. This may impede dividing assets and liabilities into very detailed accounts, such as buildings, trade names, inventory, accounts receivable, marketable securities, accounts payable, short-term loans, bonds, long-term loans, capital leases, and so on, some of which are zero for a large portion of firms, especially if the sample contains small and medium enterprises.

In other words, the choice of the number and detail of the $D$ accounting figures has to be subject to the presence of zeros. If some accounting figures contain more than 20 % zeros, the user may want to sum them with other conceptually similar accounting figures with fewer zeros and thus reduce $D$. For instance, if short-term loans have 30 % of zeros and accounts payable have 5 % of zeros, summing both into a current liabilities category will result in at most 5 % of zeros (or less if zeros do not co-occur for the same firms). Before deciding which accounting figures to aggregate it is therefore useful to examine not only their percentages of zeros and their conceptual similarity, but also zero co-occurrence by means of the so-called *zero patterns plot* which displays the frequencies of all possible combinations of zeros. In the previous example, if no firm has zeros for both short-term loans and accounts payable, the aggregated figure will be completely free of zeros. These aggregations are called amalgamations in the CoDa literature.

Related to the zero problem, inactive firms, as revealed by having zero revenues and/or zero assets should be removed from the dataset. If firms are inactive, they just do not belong to the study population, and it makes no sense to replace their missing accounting information with any sort of meaningful small value (just imagine what margin or turnover would look like with revenues or assets replaced with very small values). We recommend researchers to always drop these firms, both from a conceptual and a practical point of view. This situation is called indistinctively *essential zeros*, *structural zeros*, *absolute zeros* or *true zeros* in the CoDa literature, and consensus is that they are not fit for replacement (Martín Fernández et al., 2011).

## 4. Example data

The financial statements in this example were obtained from the SABI (Iberian Balance sheet Analysis System, accessible at https://sabi.bvdinfo.com/) database, developed by *INFORMA D&B* in collaboration with *Bureau Van Dijk*. Search criteria were wine producers in Spain (NACE Rev.2 classification code 11.02 “manufacture of wine from grape”) with available data for 2016. Inactive firms with zero revenues and/or zero assets were removed from the dataset. The final sample size after filtering inactive firms was $n$=109 and there were no remaining zero values. When using few and broad account categories as in DuPont analysis, zeros are rarely a problem.

In addition to $x_1$ to $x_4$, we consider a non-financial indicator: the qualitative variable indicating if the firm sells at least some products using its own brand (*own brand*: 1=yes, 0=no). This indicator is of especial interest, since it reflects two winery business models which have deep implications. Firms without brands tend to sell non-bottled young wines at lower prices, while branded wines tend to be aged and expensive. Thus, firms without brands tend to have lower margins but higher turnovers and firms with brands the opposed characteristics. They constitute two *strategic groups* pursuing high ROE values through two different means.

Firm age in years is also included as a non-financial firm characteristic. This dataset was also used in Linares-Mustarós et al. (2022), in Coenders (2025), and in Coenders and Arimany-Serrat (2025b), and is shown in Appendix 1.

All analyses were carried out with CoDaPack2.03.10 (Comas-Cufí and Thió-Henestrosa, 2011; Thió-Henestrosa and Martín-Fernández, 2005), an intuitive menu-driven freeware for CoDa developed by the *Research Group in Statistics and Compositional Data Analysis* at the University of Girona (https://ima.udg.edu/codapack/). See Ferrer-Rosell et al. (2022) for a gentle introduction to the CoDa methodology and the CoDaPack software. A guide to the menus used in this article is in Appendix 2.

The *boxplot* is an exploratory graphical display showing the division of firms into four equal-sized groups. Below the box there are 25 % of firms with the lowest values. The line dividing the box is the median, not to be mistaken with the average. The next 25 % of firms are between the lower box edge and the median, and the next 25 % between the median and the upper box edge. Above the box there are 25 % of firms with the highest values. Thus, half of the firms have values below and above the median, which represents the central firm in the sample. Also, half of the firms have values within the box boundaries and represent the most representative firms. The whiskers (vertical lines above and below the box) reach out to the last non-outlying value, outliers being identified as separate points. The overall appearance of the boxplot tells about the symmetry or lack of symmetry of the distribution, and the extreme points tell about the presence of outliers.

As reported in the literature, pairwise and centred log-ratios (Figs. 3 and 4) tend to be better behaved than standard ratios (Fig. 2) in terms of asymmetry and outliers. Standard ratios are not appropriate for statistical analysis, having strong asymmetry, extreme outliers, or both. In our example, the standard leverage ratio has one especially extreme outlier and strong asymmetry. Turnover has strong asymmetry. ROE is approximately symmetric but has also two very extreme outliers. Moderate outliers are usually not harmful to the results of statistical analysis. Extreme outliers are. Neither pairwise nor centred log-ratios exhibit strong asymmetry or any extreme outlier.

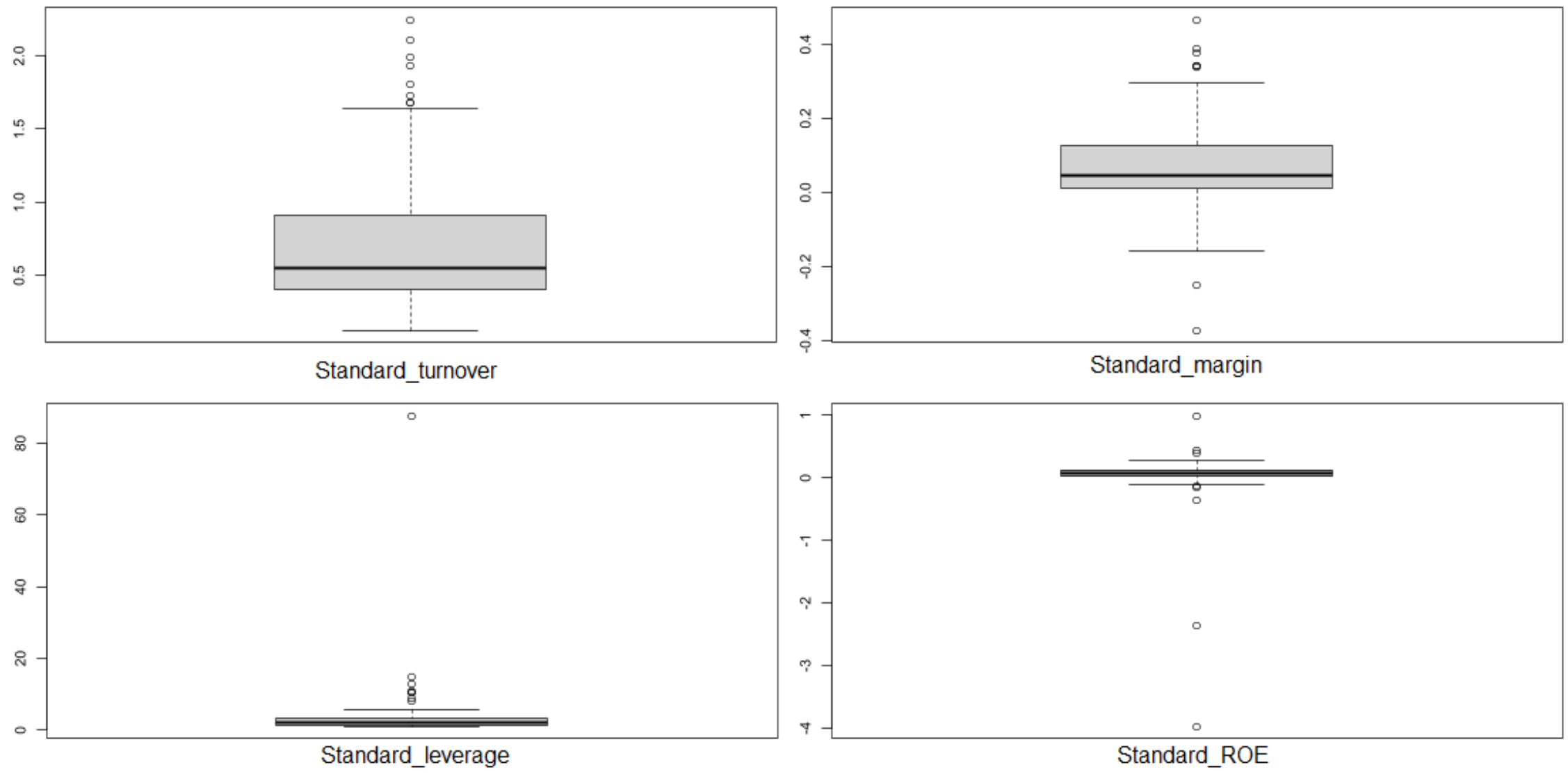

**Figure 2.** Boxplots of standard ratios in Equations (3) to (7)

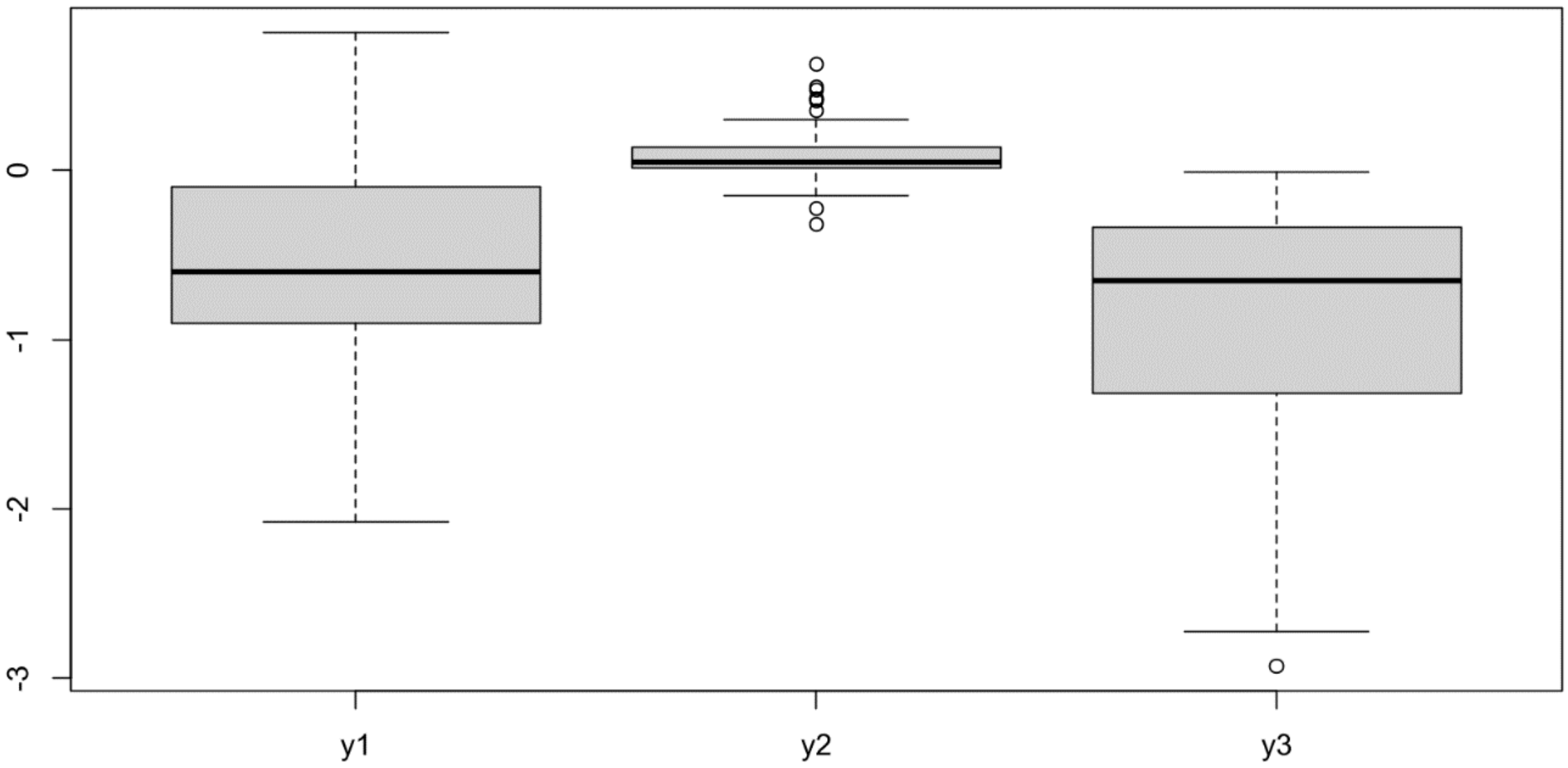

**Figure 3.** Boxplots of pairwise log-ratios in Equations (10) to (12)

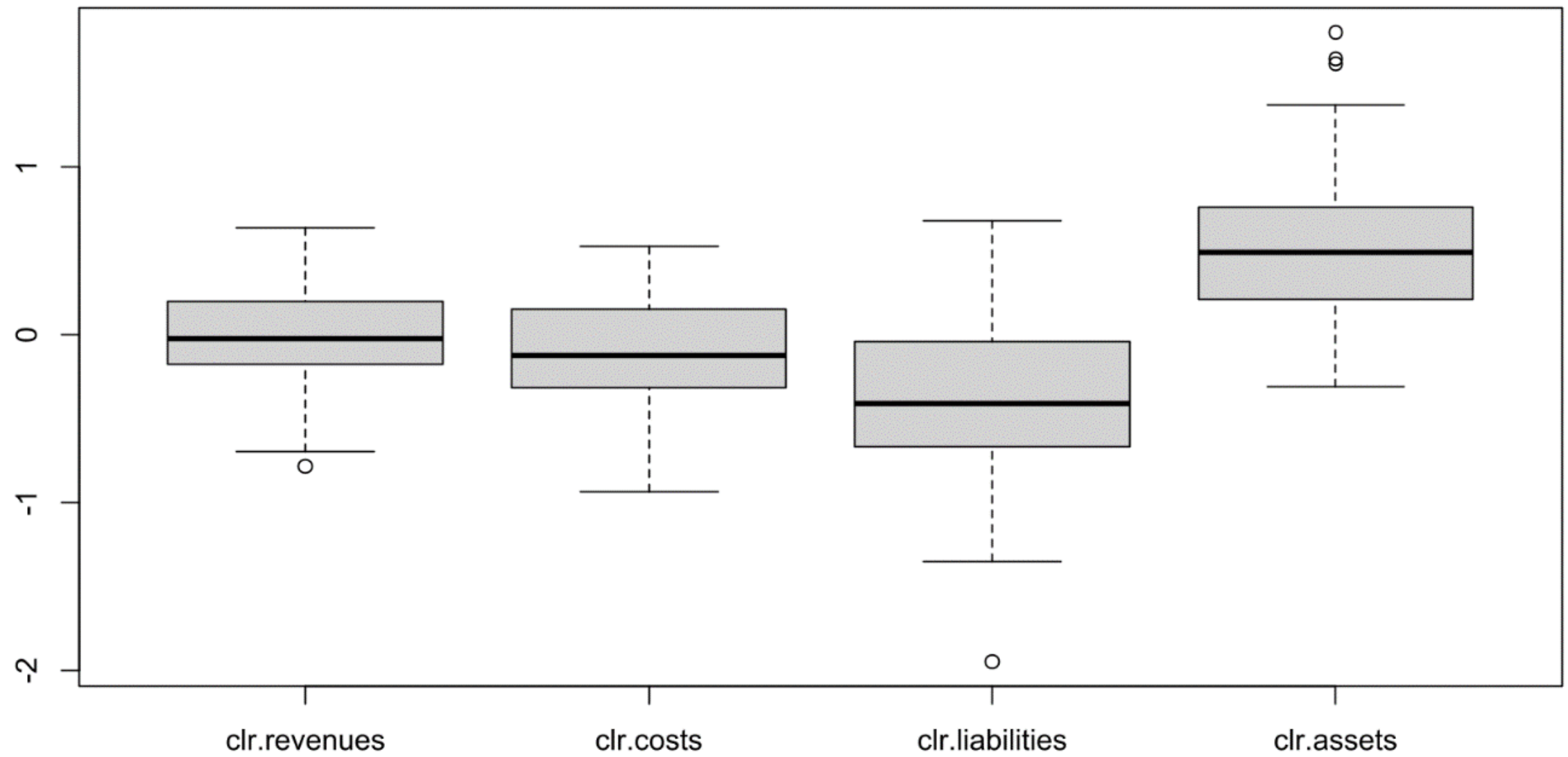

**Figure 4.** Boxplots of centred log-ratios in Equation (16) labelled according to the accounting category in the numerator

If centred log-ratios (Fig. 4) contain some remaining extreme outliers, the corresponding firms can be identified and removed from the sample. This is not the case in our example dataset.

## 5. Industry analysis I. Industry ratio averages

The simplest conceivable statistical usage of financial ratios is to compute ratio averages within an industry. The *compositional centre* (Aitchison, 1997) is defined as the array of geometric means of all firms for each individual part, normalized to unit sum for convenience (Table 2) and is used to compute the average values for compositional data:

| | |
|---|---|
| $x_1$: Revenues | 0.2354 |
| $x_2$: Costs | 0.2149 |
| $x_3$: Liabilities | 0.1590 |
| $x_4$: Assets | 0.3907 |

**Table 2.** Compositional centre (all wineries)

This is not to be mistaken with the geometric means of all parts for each individual firm used for computing the centred log-ratios in Equations (15) and (16).

Like ratios, geometric means focus on relative rather than absolute differences, are mutually compatible, and should be used together for variables in a ratio scale. If we again take firms 3, 4, and 5 in the toy example in Table 1, the geometric mean of the $x_2$ values 100, 1,000 and 10,000 is $g(x_2) = \sqrt[3]{100 \times 1{,}000 \times 10{,}000} = 1{,}000$. This is so because, in relative terms, the difference between 1,000 and 100, which is 1,000/100=10, is the same as the relative difference between 10,000 and 1,000, which is 10,000/1,000=10. Conversely, the arithmetic mean is closer to the highest absolute values disregarding the relative differences: $\overline{x}_2 = (100 + 1{,}000 + 10{,}000)/3 = 3{,}700$.

The centre computed as a geometric mean under the CoDa approach makes it possible to compute average standard financial ratios at industry level (Arimany-Serrat and Coenders, 2025; Saus-Sala et al., 2021; 2023; 2024). The geometric mean has the attractive property that the ratio of the geometric means of two parts equals the geometric mean of their ratios. Let g($x_i$) be the geometric mean of part *i* over a sample of firms:

$$g\left(\frac{x_i}{x_j}\right)=\frac{g\left(x_i\right)}{g\left(x_j\right)}. \qquad (18)$$

In the same toy example in Table 1, the geometric mean of the $x_2/x_1$ ratios for firms 3, 4, and 5 is $g\left(x_2/x_1\right)=\sqrt[3]{0.01\times 1\times 100}=1$, which is equal to the ratio of the geometric means of $x_2$ and $x_1$ $g\left(x_2\right)/g\left(x_1\right)=1{,}000/1{,}000=1$.

The arithmetic mean does not have this property. Computing first arithmetic means of the accounting figures at industry level and then standard financial ratios between those means may stand in contradiction with the results of computing first standard ratios at firm level and then the ratio arithmetic means (Saus-Sala et al., 2021).

In the same toy example in Table 1, the arithmetic mean of the $x_2/x_1$ ratios for firms 3, 4, and 5 is $\left(0.01+1+100\right)/3=33.67$ which is not the ratio of the arithmetic means of $x_1$ and $x_2$ $\bar{x}_1 / \bar{x}_2 = 3{,}700 / 3{,}700 = 1$ .

Geometric means have another appealing property in financial statement analysis. The geometric mean of a permuted ratio is the inverse of the geometric mean of the original ratio (Arimany-Serrat and Coenders, 2025; Arimany-Serrat and Sgorla, 2024):

$$g\left(\frac{x_i}{x_j}\right)=\frac{1}{g\left(\frac{x_j}{x_i}\right)}. \qquad (19)$$

This property guarantees consistency of results of two researchers using permuted versions of the same ratio. In the same toy example in Table 1, the geometric mean of the $x_2/x_1$ ratios for firms 4, 5, and 6 is $g\left(x_2/x_1\right)=\sqrt[3]{1\times 100\times 10{,}000}=100$ which is the inverse of the geometric mean of the $x_1/x_2$ ratios $g\left(x_1/x_2\right)=\sqrt[3]{1\times 0.01\times 0.0001}=0.01$.

The arithmetic mean does not have this property. In the same toy example in Table 1, the arithmetic mean of the $x_2/x_1$ ratios for firms 4, 5, and 6 is $\left(1+100+10{,}000\right)/3=3{,}367$ which is not the inverse of the arithmetic mean of the $x_1/x_2$ ratios $\left(1+0.01+0.0001\right)/3=0.3367$. The first result suggests $x_2$ to exceed $x_1$ by a factor of about three thousand while the second result suggests $x_1$ to be under $x_2$ by a factor about one third.

Using these properties, the industry average standard turnover ratio ($x_1/x_4$) can be computed from Table 2 as:

$$g(x_1/x_4)=g(x_1)/g(x_4)=0.2354/0.3907=0.603. \quad (20)$$

In the same vein, the average standard margin ratio is:

$$(g(x_1)-g(x_2))/g(x_1)=(0.2354-0.2149)/0.2354=0.087, \quad (21)$$

the average standard leverage ratio is:

$$g(x_4)/(g(x_4)-g(x_3))=0.3907/(0.3907-0.1590)=1.686, \quad (22)$$

and the average ROE is:

$$(g(x_1)-g(x_2))/(g(x_4)-g(x_3))=(0.2354-0.2149)/(0.3907-0.1590)=0.089. \quad (23)$$

This makes it possible to present the results of compositional industry analysis in terms of standard financial ratios, which are better understood by the accounting community than the CoDa log-ratios. The analysis may be repeated by previously identified subdivisions within the industry, for instance firms having or not at least one brand of their own (Table 3 and top panel of Table 4).

| | Group 0 (no) – $n$=24 | Group 1 (yes) – $n$=85 |
|---|---|---|
| $x_1$: Revenues | 0.2684 | 0.2259 |
| $x_2$: Costs | 0.2522 | 0.2045 |
| $x_3$: Liabilities | 0.1558 | 0.1593 |
| $x_4$: Assets | 0.3237 | 0.4102 |

**Table 3.** Compositional centre of wineries with (1) and without (0) their own brand

| | Turnover | Margin | Leverage | ROE |
|---|---|---|---|---|
| No brand (0) | 0.829 | 0.060 | 1.928 | 0.096 |
| Brand (1) | 0.551 | 0.095 | 1.635 | 0.085 |
| Overall | 0.603 | 0.087 | 1.686 | 0.089 |
| No brand (0) | 0.734 | 0.063 | 1.698 | 0.079 |
| Brand (1) | 0.581 | 0.077 | 1.836 | 0.082 |
| Overall | 0.604 | 0.075 | 1.814 | 0.082 |
| No brand (0) | 1.030 | 0.055 | 4.309 | 0.110 |
| Brand (1) | 0.644 | 0.086 | 3.518 | 0.002 |
| Overall | 0.729 | 0.079 | 3.692 | 0.026 |

**Table 4.** Top panel: Standard ratios of wineries with (1) and without (0) their own brand computed from the geometric means in Tables 2 and 3 as in Equations (20) to (23). Centre panel: Standard ratios computed from the arithmetic means of $x_1$ to $x_4$. Bottom panel: Arithmetic means of the standard ratios

If we look at the top panel of Table 4, as expected, firms with no brands have lower margin and higher turnover. For instance, the average turnover for firms with no brands can be computed from the geometric means in Table 3 as 0.2684/0.3237=0.829 and for firms with brands as 0.2259/0.4102=0.551. Firms with no brands also have higher leverage. All in all, the ROE is more favourable for firms with no brands, at the expense of a higher indebtedness.

The centre and the bottom panels of Table 4 illustrate what happens when using the industry arithmetic means of the accounting figures to compute the standard industrial ratios, or when using the arithmetic means of the standard ratios at firm level, respectively. Changes can be dramatic. For instance, the average ROE of firms with a brand looks much lower in the bottom panel. The average leverage looks much higher in the bottom panel for all kinds of companies. In the centre panel, firms with no brand have the lowest leverage; in the bottom panel, the highest. We recommend always to use the geometric-mean approach.

Up to here, we have learned that when the data are in a ratio scale, meaning that relative and not absolute differences are of interest, ratios, logarithms and geometric means constitute meaningful operations that should be used together. It makes no sense to use ratios pretending that relative differences are being sought and then fail to use the logarithm or fail to use the geometric mean as if absolute differences had been sought. These three operations are the core of CoDa analysis.

As a footnote to this section, one may wonder why industry averages are not computed from log-ratios. Implicitly they are. It can be proven that the arithmetic means computed on the centred log-ratios are equivalent to the geometric means computed from the accounting figures that have been presented here. The only thing which needs to be done is to exponentiate the arithmetic clr means (Aitchison, 1997). In this article we use the geometric mean representation due to its intuitive appeal.

## 6. Industry analysis II. Visualisation of individual firms with the CoDa biplot

Like any other statistical data, compositional data require visualization tools to help researchers interpret large data tables with many firms and parts. To this end, Aitchison (1983) extended the well-known *principal component analysis* procedure (Hotelling, 1933; Greenacre et al., 2022) to the compositional case. This method belongs to the family of *multivariate statistical methods*, and the extension boils down to submitting the covariance matrix of the $D$ centred log-ratios in Equation (16) to a principal component analysis.

A compositional principal component analysis computes a small number of uncorrelated linear combinations of the centred log-ratios, called dimensions, which explain the highest possible portion of the sum of the variances of all centred log-ratios. In this way the original data set with potentially many centred log-ratios can be summarized with just a few dimensions which are suitable for a graphical display.

The two first dimensions are represented in the so-called covariance *CoDa biplot* (Aitchison and Greenacre, 2002, drawing from Gabriel, 1971), which can be understood as the most accurate graphical representation of a compositional data set in two dimensions. The goodness of fit is indicated by the percentage of explained variance of the centred log-ratios by the first two dimensions. In our example, the percentage of explained variance by the first two dimensions is very high at 98.99 % thus arguing for an extremely good biplot accuracy. The information in the original data can be represented in a two-dimensional biplot with very high precision. From our own experience, percentages above 70 % can be considered acceptable, percentages above 80 % good and percentages above 90 % very good.

The CoDa biplot for financial statement data plots each centred log-ratio representing the accounting figure in the numerator as a line called *ray*. Individual firms appear as points.

Carreras-Simó and Coenders (2020); Rondós-Casas et al. (2026); and Saus-Sala et al. (2021; 2023) highlight the most important interpretational tool of the CoDa biplot in financial statement analysis. Additional lines can be drawn linking the extremes of a pair of rays and representing the pairwise log-ratios between the corresponding two accounting figures of the numerators of the centred log-ratios. These additional lines are called *links*. The *orthogonal projection* of all firms along the direction defined by the link between the vertices of a pair of rays shows an approximate ordering of firms according to the pairwise log-ratio between the corresponding two accounting figures. The orthogonal projection is made by dropping firms on the link in such a way that the direction in which the firms fall forms a 90-degree angle with the link.

In this way, the CoDa biplot is also a visual representation of any of the $D(D–1)/2$ possible financial ratios computed from any two accounting figures. The user can draw as many links as he or she wishes. Since the analysis is anyway run on centred log-ratios, redundancy is not a problem, although only long links showing high variance pairwise log-ratios tend to lead to informative directions. Thus, pairwise ratios should not be drawn when the links are very short, in other words, when the vertices of the two involved centred-log-ratio rays are close together.

In our case, the three pairwise log-ratios of interest are $y_1$ (turnover), $y_2$ (margin) and $y_3$ (leverage) and have thus been drawn above the biplot (Fig. 5). Since ROE cannot be expressed as a pairwise ratio, it cannot be represented in the biplot.

The ability to visually interpret ratios between any two accounting figures is of great interest in financial statement analysis in general and in DuPont analysis in particular (Saus-Sala et al., 2021; 2023). The $y_1$ line representing turnover links the vertices of revenues and assets. Firms situated at the top and to the right (in the high-revenue side of the link) are the ones with the highest turnover. Firms situated at the bottom and to the left (in the high-asset side of the link) are the ones with the lowest turnover. The $y_2$ line representing margin links the vertices of revenues and costs. Firms situated at the bottom and to the left are the ones with the highest margin and firms situated at the top and to the right are the ones with the lowest margin. The $y_3$ line representing leverage links liabilities and assets. Firms at the top of the biplot are the most leveraged.

In more precise terms, firms are dropped forming a 90-degree angle on each of the $y_1$, $y_2$ and $y_3$ lines. For instance, firm 16 has the highest orthogonal projection on $y_2$ and the lowest on $y_3$. Thus, it is a firm with a very high margin and a very low leverage. Firm 59 has a very low turnover and firm 52 has a very high turnover. As a whole, firms without a brand (marked as grey) have a comparatively lower margin, higher leverage and higher turnover than firms with a brand (marked as black). The closest firms to the origin of coordinates, like firms 8, 17, 41, 71, 81 and 91, are also the closest to the industry average described in Section 5.

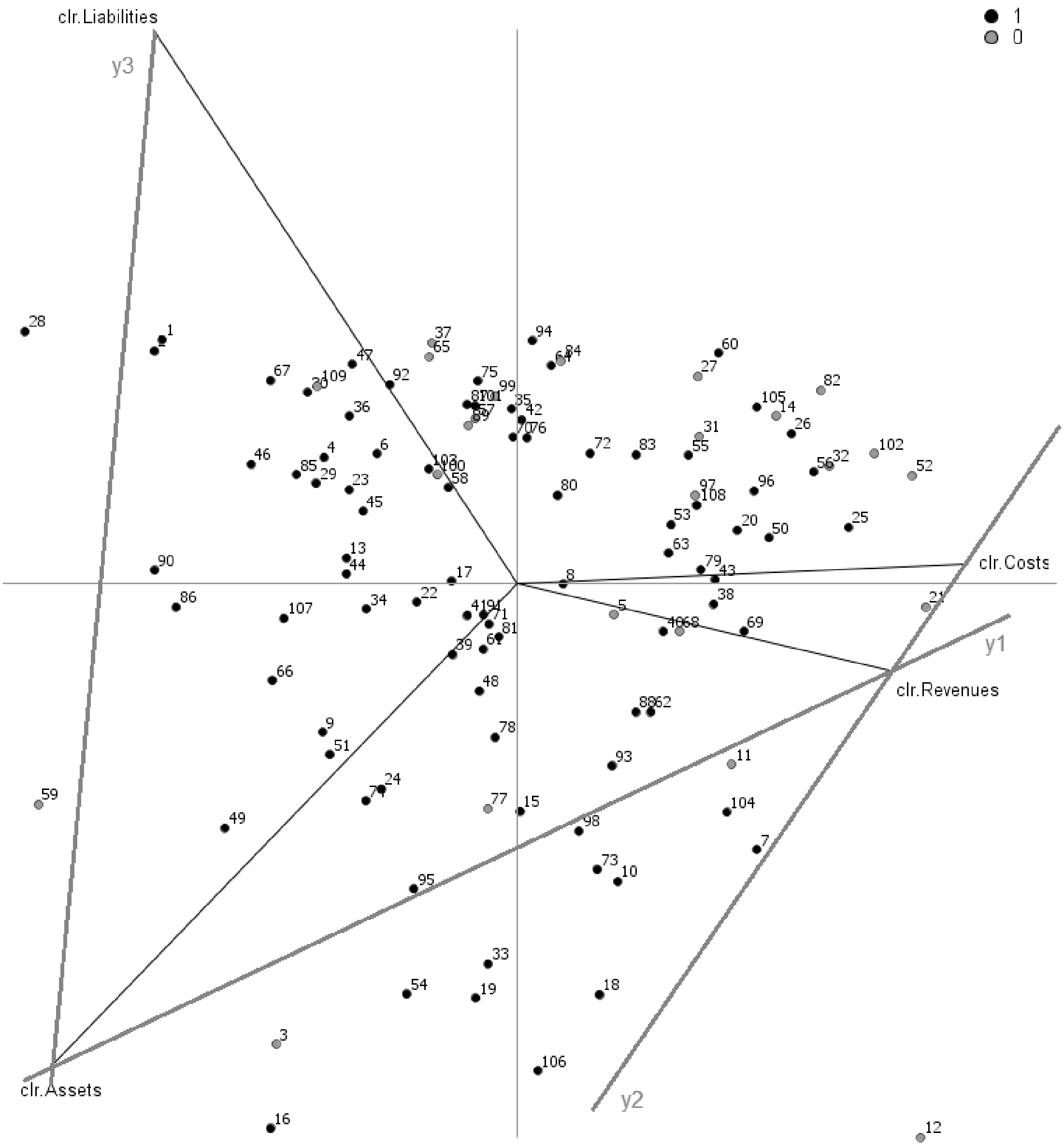

**Figure 5.** CoDa biplot with added turnover (y1) margin (y2) and leverage (y3) directions. Firms with an own brand in black, without in grey

The compositional biplot thus becomes an intuitive and useful tool for strategic analysis (Carreras-Simó and Coenders, 2020) as it allows researchers to quickly identify individual firms competing on the basis of margin, on the basis of turnover, or on the basis of leverage.

## 7. Industry analysis III. Industry heterogeneity assessment with cluster analysis

Very rarely can an industry be assumed to represent one single homogeneous financial-statement pattern. *Cluster analysis* is another popular multivariate statistical method which aims to extracting groups or clusters of individuals (i.e., firms) in such a way that individuals of the same group are as similar (homogeneous) as possible according to the variables of interest. In other words, firms in the same group must have low mutual distances. Likewise, firms in the different groups must be as dissimilar as possible, i.e., have large mutual distances (Kaufman and Rousseeuw, 1990). Compositional cluster analysis boils down to performing an otherwise standard cluster analysis on the $D$ centred log-ratios in Equation (16) (Ferrer-Rosell and Coenders, 2018; Martín-Fernández et al., 1998).

If centred log-ratios are used as data, Euclidean distances become equal to the standard Aitchison's distances used in CoDa (Aitchison, 1983; Aitchison et al., 2000). The distance between firms $m$ and $l$ is thus computed from the differences in their respective centred log-ratios as:

$$d_{ml} = \sqrt{\left(clr_{1m} - clr_{1l}\right)^2 + \left(clr_{2m} - clr_{2l}\right)^2 + \cdots + \left(clr_{Dm} - clr_{Dl}\right)^2} \ . \quad (24)$$

Any standard clustering method handling Euclidean distances can be used. This includes, among others, two popular clustering methods in financial-statement analysis (Linares-Mustarós et al., 2018): Ward's method (Ward, 1963), and the $k$-means method (MacQueen, 1967).

In the financial-statement context, compositional cluster analysis can be used to identify subgroups of firms with similar financial-statement structures within an industry (Arimany-Serrat and Coenders, 2025; Arimany-Serrat and Sgorla, 2024; Coenders, 2025; Dao et al., 2026; Hernandez Romero and Coenders, 2025; Jofre-Campuzano and Coenders, 2022; Linares-Mustarós et al., 2018; Molas-Colomer et al., 2024; 2026; Saus-Sala et al., 2021; 2023; 2024). This has sometimes been called "profiling financial performance and financial distress".

In the example we use the $k$-means method. To classify the firms into $k$ clusters, this method:

- takes $k$ random firms as initial cluster centres,
- each of the remaining firms is assigned to the cluster with the closest centre (i.e., whose centre is at the lowest Euclidean/Aitchison distance),
- the centres are recomputed as the clr arithmetic means of the firms in each cluster.

The reassignment of firms and update of the centres is repeated until no firm moves cluster between one step and the next. Since the final result may depend on which firms are taken as initial centres, the process is repeated several times with different initial cluster centres randomly chosen. CoDaPack performs such 25 repetitions. Only the solution with the highest cluster homogeneity (lowest sum of clr variances within the clusters) is presented to the user.

The appropriate number of clusters is rarely known in advance. Several statistical criteria are available to decide the best $k$ after doing classifications with reasonable numbers of clusters, for instance from $k$=2 to $k$=8. In our example, a 3-cluster solution maximises both the *average silhouette width* (Kaufman and Rousseeuw, 1990) at 0.422, and the *Caliński-Harabasz index* (Caliński and Harabasz, 1974) at 86.9. The number of clusters can also be chosen according to interpretability: adding a cluster makes sense if it adds a meaningfully different financial-statement profile, without leading to any of the existing clusters being very small. One starts with 2 clusters and keeps on adding clusters as long as the above statement holds. From our own experience, the ideal number of clusters is usually between 3 and 5. In a two-cluster solution one cluster merely has opposite characteristics from the other in all ratios, which is rather uninteresting. A solution with more than 5 clusters tends to be very hard to interpret.

From the cluster compositional centres (i.e., the geometric means), the standard financial ratios in Equations (3) to (7) can be computed to represent an average firm in the cluster (Tables 5 and 6), which makes for a simple interpretation. Cluster 1 (36 firms) has the highest turnover and ROE and the lowest margin. Cluster 2 (23 firms) has the lowest turnover, leverage and ROE and the highest margin, and Cluster 3 (50 firms) the highest leverage. The practitioner can compute as many standard financial ratios as he or she wishes. Since the analysis is anyway run on centred log-ratios, the redundancy of ROE with respect to turnover, margin, and leverage is not a problem.

| | Cluster 1 – $n$=36 | Cluster 2 – $n$=23 | Cluster 3 – $n$=50 |
|---|---|---|---|
| $x_1$: Revenues | 0.3090 | 0.1923 | 0.1934 |
| $x_2$: Costs | 0.2979 | 0.1549 | 0.1797 |
| $x_3$: Liabilities | 0.1324 | 0.0788 | 0.2281 |
| $x_4$: Assets | 0.2607 | 0.5739 | 0.3988 |

**Table 5.** Compositional centre of wineries per cluster

| Cluster | Turnover | Margin | Leverage | ROE |
|---|---|---|---|---|
| 1 | 1.185 | 0.036 | 2.032 | 0.087 |
| 2 | 0.335 | 0.194 | 1.159 | 0.076 |
| 3 | 0.485 | 0.071 | 2.336 | 0.080 |

**Table 6.** Standard ratios of wineries computed from the cluster geometric means as in Equations (20) to (23)

The situation in the biplot in reference to the directions defined by the pairwise log-ratios $y_1$, $y_2$ and $y_3$ is a further interpretational aid. The biplot can be redrawn with the firms coloured by the cluster membership variable (Fig. 6). From Fig. 5 it must be recalled that firms with the highest turnover ($y_1$) are situated at the top right of the graph, firms with the highest margin ($y_2$) at the bottom left, and firms with the highest leverage ($y_3$) at the top.

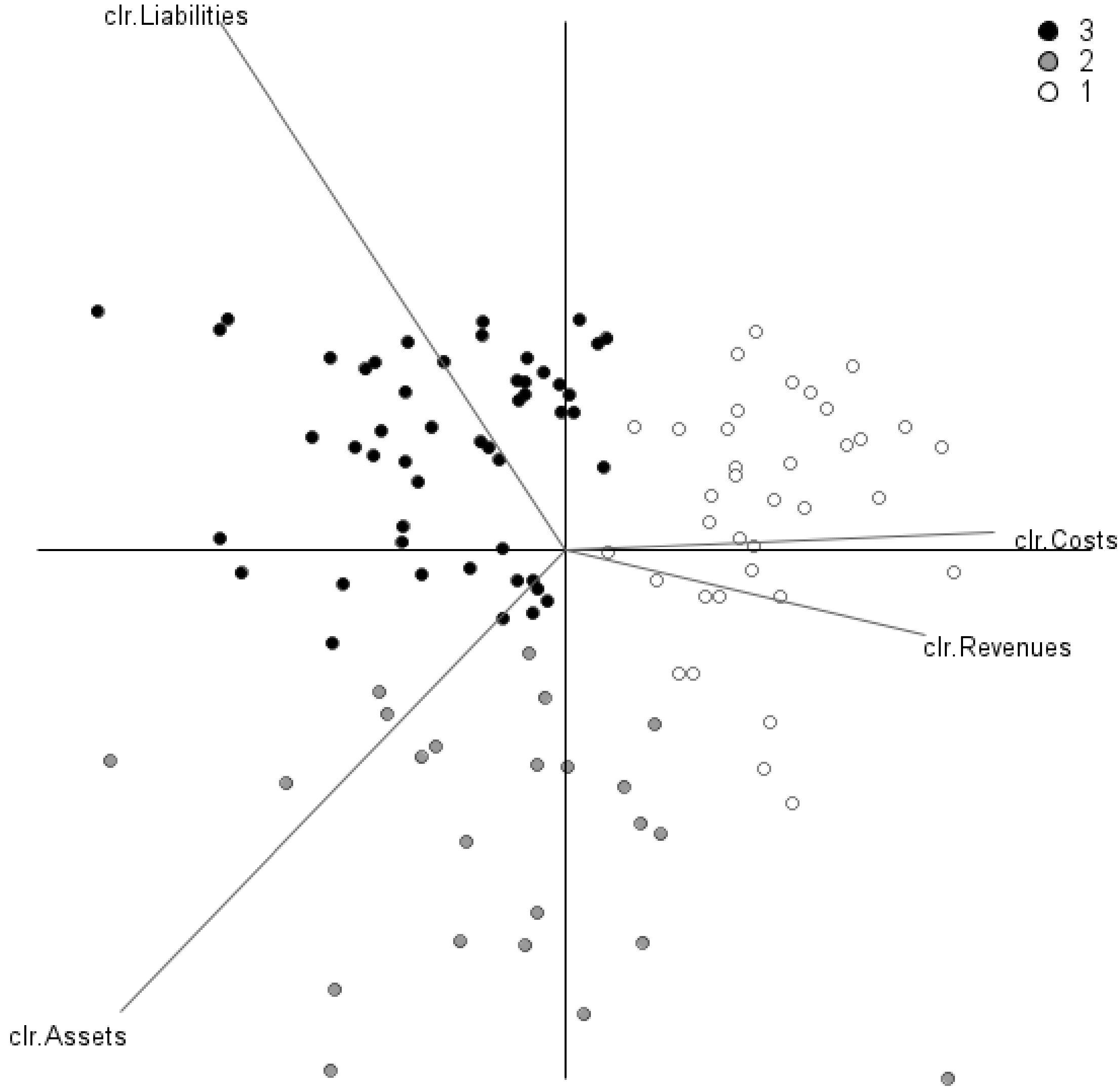

**Figure 6.** Compositional biplot with firm points coloured by cluster membership

Mosaic plots can be used to relate the cluster membership to *categoric* (a.k.a. *qualitative*) non-financial indicators and firm characteristics, like having or not having an own brand (Dolnicar et al., 2018). Fig. 7. shows Clusters 2 and 3 to be more prevalent in wineries with an own brand (1), and Cluster 1 in firms without any own brand (0). This makes theoretical sense as Cluster 1 has the highest turnover and the lowest margin.

Boxplots can be used to relate the cluster membership to numeric non-financial indicators and firm characteristics such as firm age. Fig. 8 shows the median age to be lower for Cluster 1. All things taken together, Cluster 1 shows a very distinct profile, with the lowest age, the largest share of firms without brand, the highest turnover, and the lowest margin.

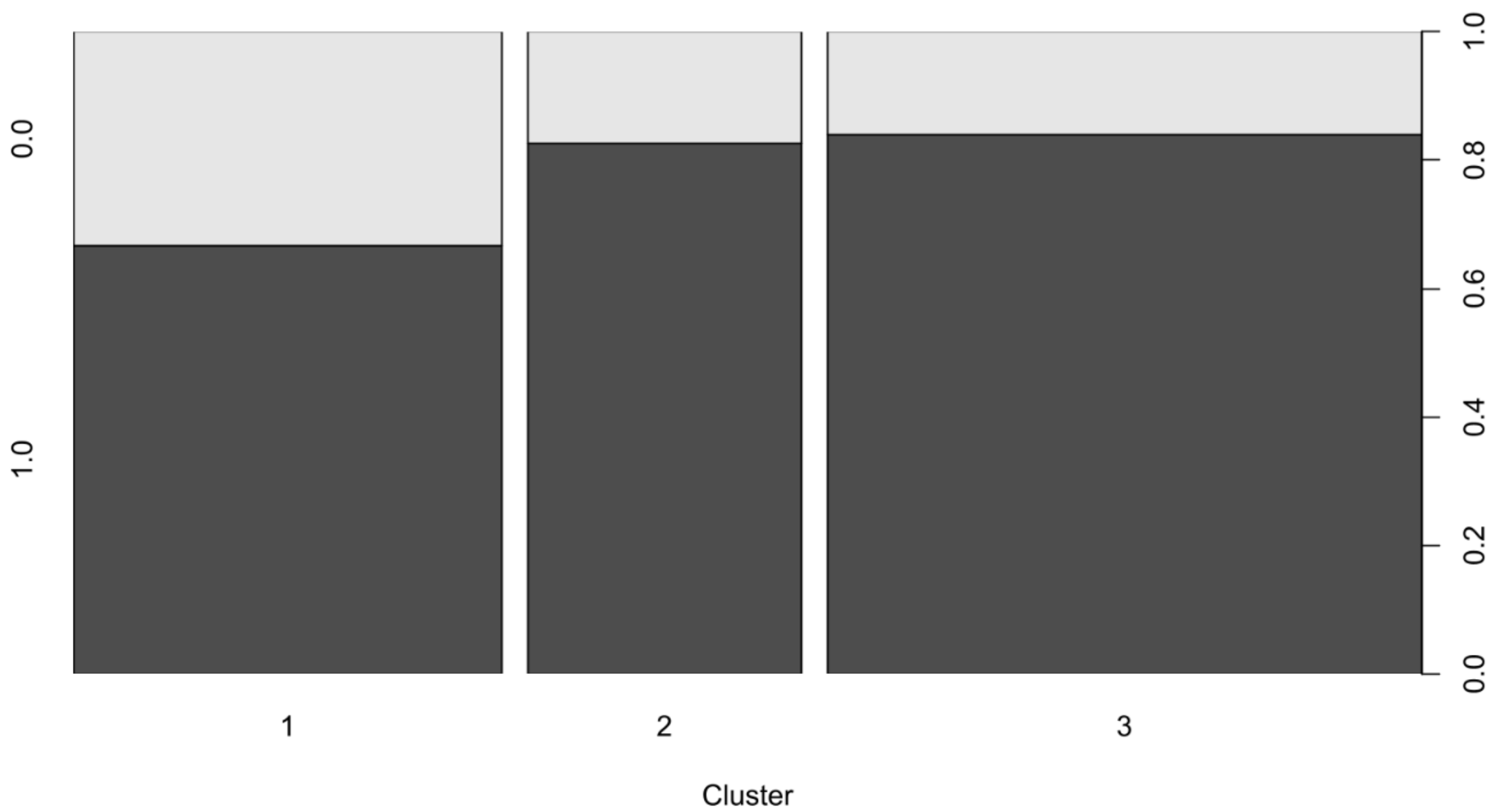

**Figure 7.** Mosaic plot of cluster and having (1) or not (0) an own brand. Bar heights are percentages of firms with and without brand within a cluster. Bar widths are cluster sizes. Bar areas are firm counts within each of the cluster-brand combinations

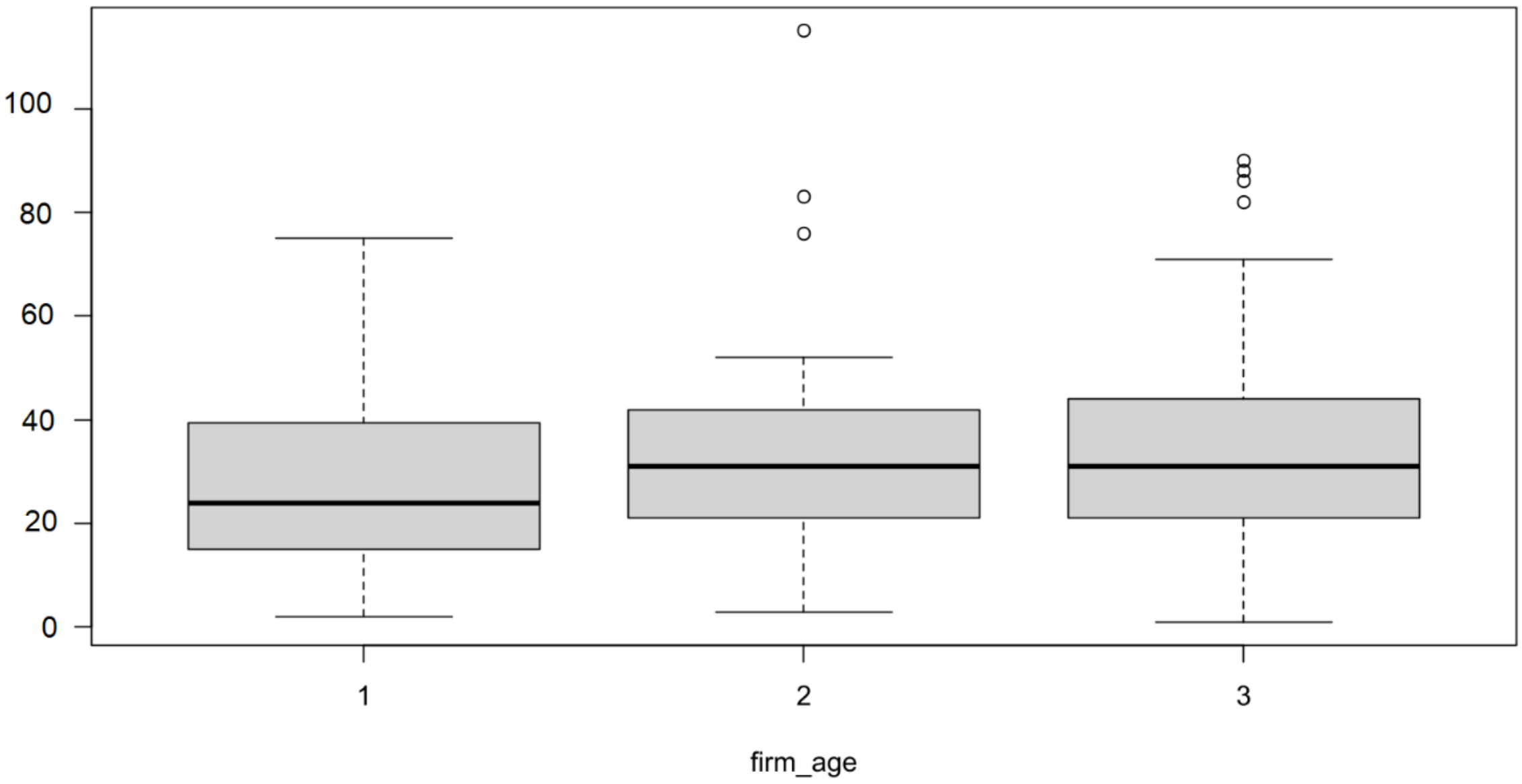

**Figure 8.** Boxplots of firm age in years by cluster

We also examined the clustering solutions with 4 and 5 clusters (results not shown). The solution with 4 clusters basically preserved clusters 1 and 2 in the 3-cluster solution while subdividing cluster 3 into two new clusters differing mainly in turnover. The 4-cluster solution would also have made for an interesting interpretation. In the 5-cluster solution, one of the clusters was very small, thus rendering the classification less useful.

# 8. Industry analysis IV. Relationships to non-financial indicators and firm characteristics

Up to now we have dealt with descriptive statistical methods. This section is devoted to statistical modelling, inference, and testing. To this end, compositional financial ratios act as variables in any statistical model together with non-financial indicators, firm characteristics, characteristics of the entrepreneur, managerial styles, etc.

Centred log-ratios are recommended for multivariate descriptive statistical analyses (e.g., cluster analysis, biplot, and principal component analysis as used in Sections 6 and 7) but not for certain types of statistical models, for which alternative log-ratio transformations are preferrable. Even of more practical importance, centred log-ratios are not directly interpretable as variables in accounting, while in statistical models the interpretation of the included variables is a crucial issue, which makes a set of $D-1$ pairwise log-ratios a preferable option. As indicated in Section 3.1, in order to include the whole information in the $D$ parts while avoiding redundancy, pairwise log-ratios must form a connected acyclic graph. $y_1$ to $y_3$ according to Equations (10) to (12) are a feasible choice. An alternative is presented in Appendix 3.

## 8.1. Compositional financial ratios as dependent variables

We first consider the case in which the log-ratios play the role of *dependent variables a.k.a. predicted or explained variables*; the reverse case is in Section 8.2. Once suitable log-ratios have been computed, a statistical model can be performed with standard methods, starting with *ordinary-least-squares linear regression* in which the composition (i.e., the transformed financial ratios) is made to dependent on one or more non-compositional *independent variables a.k.a. predictor or explanatory variables*. The statistical concepts are developed in Aitchison (1982); Egozcue et al. (2012); and Tolosana-Delgado and Van den Boogaart (2011). Applications to financial statements are in Arimany-Serrat et al. (2023); Coenders (2025); Escaramís and Arbussà (in press); Farreras-Noguer et al. (2025); and Mulet-Forteza et al. (in press). The predictors may not only be numeric but also qualitative with two categories (i.e. binary), as long as the two categories are coded as 0 and 1. This makes it possible to predict the financial indicators contained in the financial-statement composition from non-financial indicators and other firm or management characteristics. The reader unfamiliar with ordinary-least-squares linear regression and with statistical hypothesis testing is advised to resort to any introductory statistics or econometrics handbook.

Before modelling, some graphical display relating the log-ratios with the non-financial indicators and firm characteristics is very useful. As in the previous section, we consider the brand variable and firm age. According to the boxplots (Fig. 9), at first sight, firms with a brand have higher margin ($y_2$) but lower turnover ($y_1$) and lower leverage ($y_3$). There are far fewer outliers and far less asymmetry than in Fig. 2. According to the *scatterplots* (Fig. 10), at first sight there is little or no relationship between firm age and any of the pairwise log-ratios, and there might be an outlier in firm age, with a firm aged over 100 years.

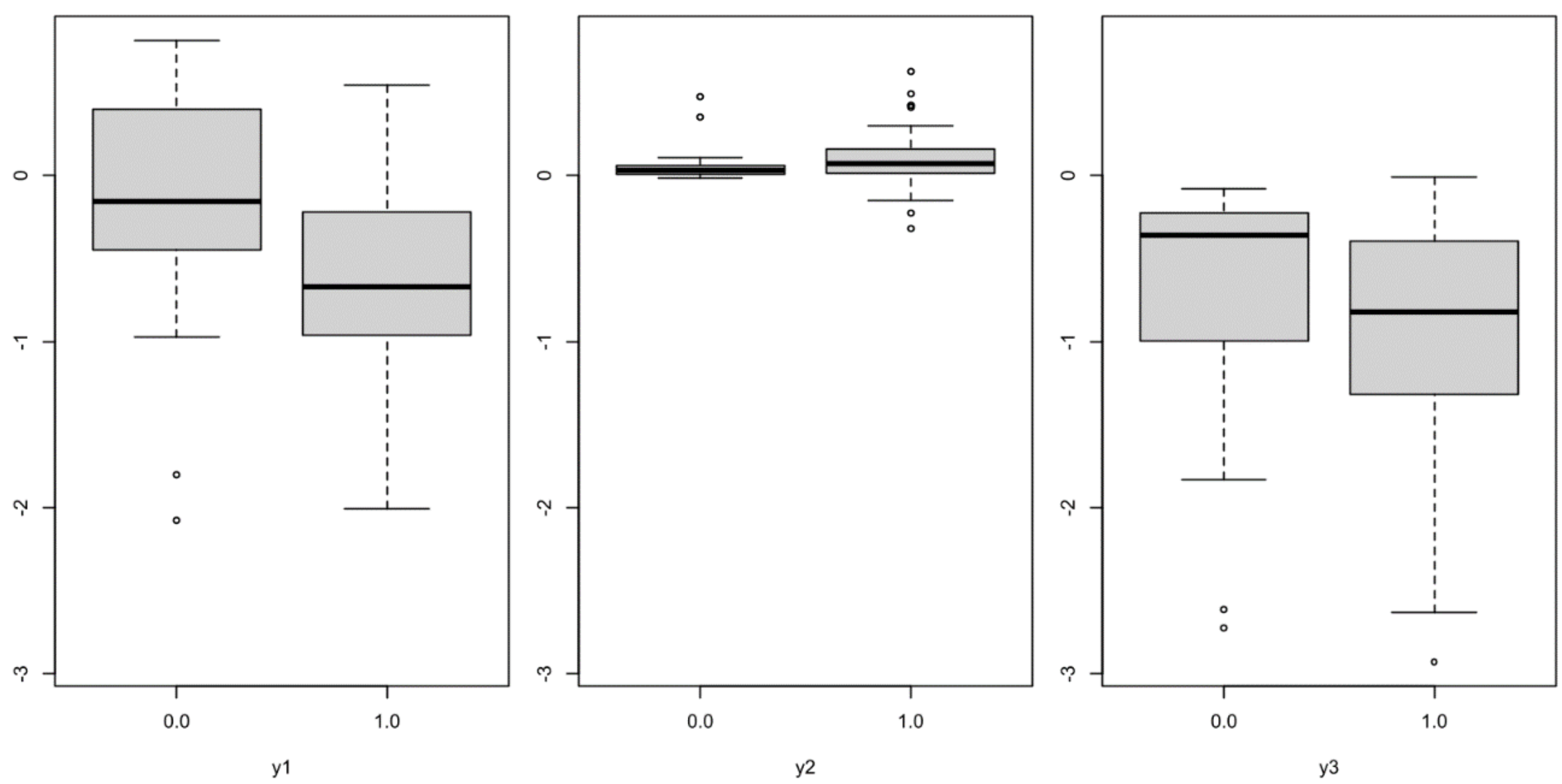

**Figure 9.** Boxplots of pairwise log-ratios in Equations (10) to (12) for wineries with (1) and without (0) their own brand

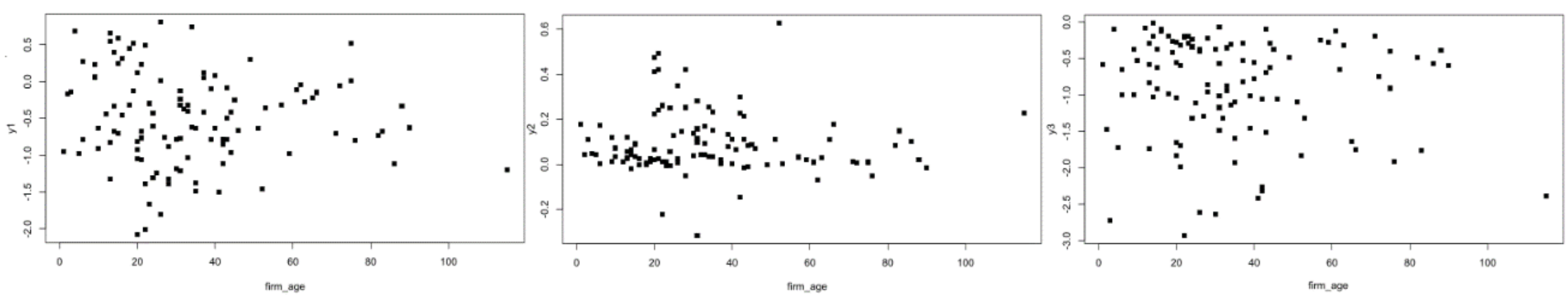

**Figure 10.** Scatterplots of pairwise log-ratios in Equations (10) to (12) with firm age in years

In our example we note the predictor firm age in years as $z_1$ and the non-financial indicator distinguishing firms with a brand as $z_2$. The qualitative variable $z_2$ is appropriately coded as 0 (no brand) and 1 (at least one brand). The $y_1,\ldots y_{D-1}$ pairwise log-ratios are the dependent variables in $D$–1 linear regression equations specified as:

$$
\begin{aligned}
y_1 &= \alpha_1 + \beta_{11} z_1 + \beta_{12} z_2 + \varepsilon_1 \\
y_2 &= \alpha_2 + \beta_{21} z_1 + \beta_{22} z_2 + \varepsilon_2 \\
y_3 &= \alpha_3 + \beta_{31} z_1 + \beta_{32} z_2 + \varepsilon_3,
\end{aligned}
\qquad (25)
$$

where $y_1$ to $y_3$ are the pairwise log-ratios in in Equations (10) to (12), $z_1$ and $z_2$ are the predictor variables, the $\alpha$ parameters are the intercept terms, and the $\beta$ parameters are effects of each of the $z$ predictors on each of the pairwise log-ratios. These effects are interpreted keeping all other predictors of the same log-ratio constant. The $\varepsilon$ terms are the residuals standing for the part of the pairwise log-ratios which is not explained by the predictors.

The predictors used must be the same for all pairwise log-ratios. This is so because one must consider the financial statement composition as just one vector variable with interrelated parts. For instance, $y_1$ and $y_2$ have the same numerator. It would not be conceivable that a variable belongs to the equation predicting $y_1$ and not to the equation predicting $y_2$.

It is not possible to include financial log-ratios at the right-hand side of the regression equations to predict another financial log-ratio. This is so because ratios involving the same parts of the composition (i.e., the same financial statements) are prone to *spurious* (i.e., *false*) *correlations*, a fact that was already revealed by Pearson himself at the time he was developing the correlation concept (Pearson, 1897), has long been acknowledged in the accounting literature (Lev and Sunder, 1979) and affects standard and compositional financial ratios equally.

The following statistical hypotheses are tested corresponding to the $\beta$ parameters in the regression equations (25). First there is a *global test* for each of the equations:

$H_0$: $\beta_{11}=\beta_{12}=0$ (none of the variables affects turnover),
$H_0$: $\beta_{21}=\beta_{22}=0$ (none of the variables affects margin),
$H_0$: $\beta_{31}=\beta_{32}=0$ (none of the variables affects leverage).

Then there is an *individual test* for each $\beta$ coefficient:

$H_0$: $\beta_{11}=0$ (firm age does not affect turnover),
$H_0$: $\beta_{21}=0$ (firm age does not affect margin),
$H_0$: $\beta_{31}=0$ (firm age does not affect leverage),
$H_0$: $\beta_{12}=0$ (having or not a brand does not affect turnover),
$H_0$: $\beta_{22}=0$ (having or not a brand does not affect margin),
$H_0$: $\beta_{32}=0$ (having or not a brand does not affect leverage).

The *p-value* associated to each statistical test indicates the risk involved in rejecting the hypothesis. If this is low (e.g., lower than 0.05), the hypothesis can be rejected. If the hypothesis in the global test is rejected, it leads to the conclusion that at least one of the explanatory variables is useful in predicting the log-ratio at hand. The individual tests next indicate which. If the hypothesis of an individual test is rejected it leads to the conclusion that the predictor at hand does affect the involved log-ratio, in other words that its effect is *statistically significant*. We can thus assess the statistical significance of the differences between firms having or not a brand, intuitively revealed by the top panel of Table 4 and the boxplots in Fig. 9 and do so keeping firm age constant.

In Table 7, the global p-value indicates that only turnover is significantly related to any of the predictor variables. Within this equation, the only individual p-value lower than 0.05 corresponds to the brand variable, thus telling that turnover is different depending on whether firms have or fail to have an own brand (p-value=0.0064). The negative sign of the coefficient estimate (−0.4068) indicates that firms with a brand (labelled as 1) have a lower turnover, keeping firm age constant. A positive sign would have indicated the opposite. Firm age is not significantly related to any of the log-ratios.

The coefficients can be interpreted in terms of the standard (not log-transformed) pairwise ratios by exponentiating them (Farreras-Noguer et al., 2025). Having a brand is associated to multiplying the value of the standard current-asset turnover ratio in Equation (3) (revenues/assets) by 0.666 ($e^{-0.4068}$).

| | Age ($z_1$) | | Brand ($z_2$) | | | Global |
|---|---|---|---|---|---|---|
| | $\beta$ estimate | p-value | $\beta$ estimate | p-value | $R^2$ | p-value |
| $y_1$ (turnover) | −0.0002 | 0.9538 | −0.4068 | 0.0064 | 0.0739 | 0.0171 |
| $y_2$ (margin) | −0.0005 | 0.4004 | 0.0447 | 0.1762 | 0.0194 | 0.3541 |
| $y_3$ (leverage) | −0.0019 | 0.5469 | −0.1869 | 0.2664 | 0.0198 | 0.3470 |

**Table 7.** Regression estimates for the pairwise log-ratios predicted by firm age and the variable indicating wineries with a brand of their own, coded as 1

The $R^2$ indicate the percentages of variance of each pairwise log-ratio explained by the $z$ variable(s). In our case, they are very low at 7.39 %, 1.94 % and 1.98 %, thus showing that other non-financial indicators and firm characteristics not considered here may have the lion's share in explaining the behaviour of turnover, margin, and leverage in wineries.

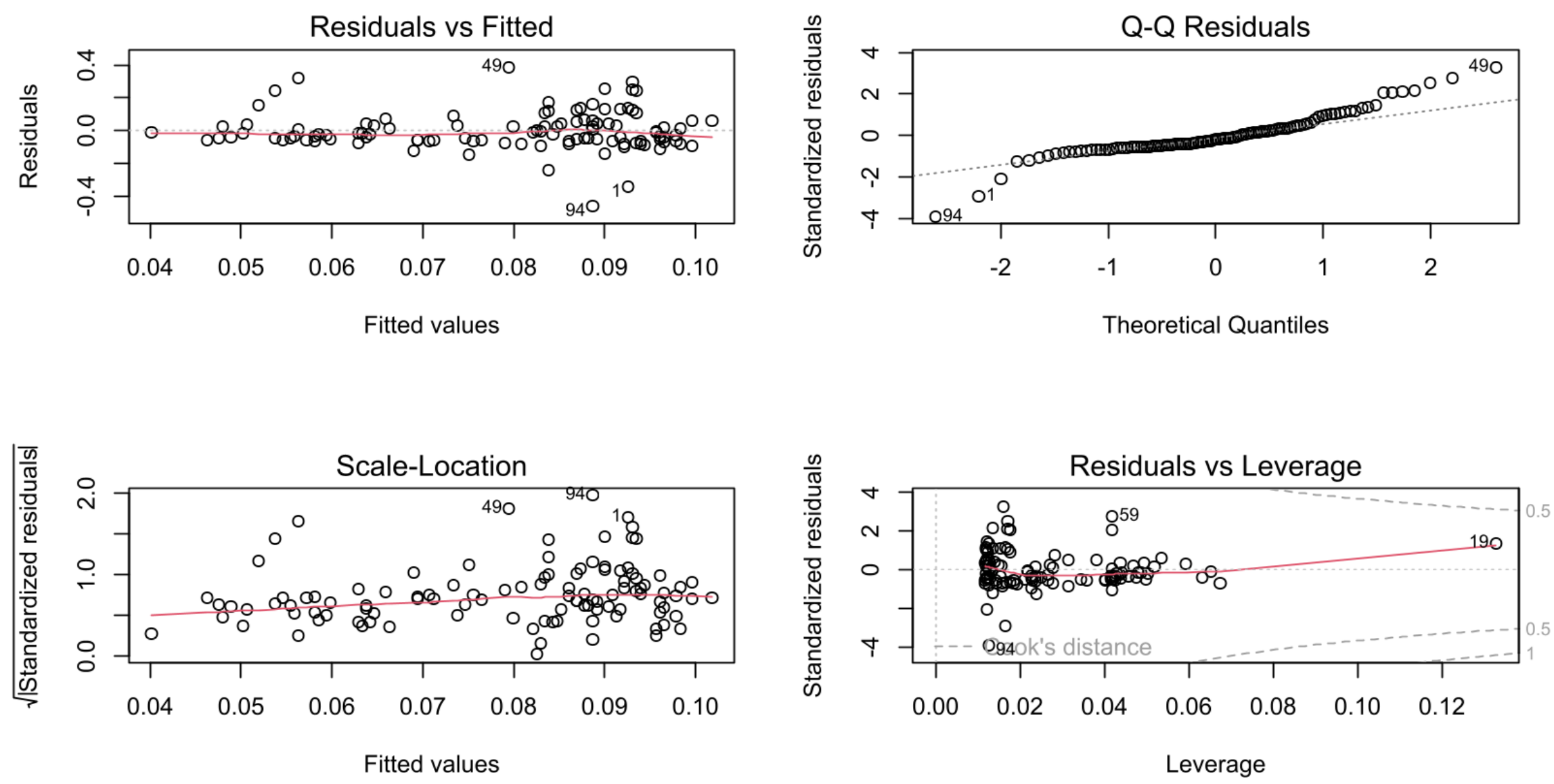

**Figure 11.** Residual plots for the equation predicting $y_2$ (margin)

Finally, for each of the three equations a set of *residual plots* is used to check the models' assumptions. Fig 11. Shows such plots for the $y_2$ (margin) equation just as an example.

- A scatterplot of the residuals versus the fitted values exhibits a linear pattern, thus showing that the *linearity* assumption is approximately met.
- A scatterplot of the square root of the absolute standardized residuals versus the fitted values exhibits a horizontal pattern with constant dispersion, thus showing that the *homoscedasticity* (a.k.a. equal variance) assumption is approximately met.
- A quantile-quantile (Q-Q) plot of the residuals exhibits a non-linear pattern, thus showing that the *normality* assumption is not met. However, violation of the normality assumption only has serious consequences for small samples.
- A scatterplot of residuals versus *leverage* is used to detect if there are *influential outliers*, which, if any, would lie at the upper right or lower right corners beyond the 0.05 border for Cook's distance. Leverage here is a statistical concept having nothing to do with financial leverage.

Readers unfamiliar with the regression model assumptions are advised to resort to an introductory statistics or econometrics handbook.

### 8.2. Compositional financial ratios as explanatory variables

An alternative possibility is that the composition (i.e., the set of transformed financial log-ratios) predicts a non-compositional numeric dependent variable *w* (Aitchison and Bacon-Shone, 1984; Coenders and Greenacre, 2023; Coenders and Pawlowsky-Glahn, 2020; Hron et al., 2012).

In this case, all log-ratios are simultaneously included at the right-hand side of one single regression equation. Additional non-financial *z* variables can also be included if they contribute to the explanatory power.

$$w = \alpha_1 + \beta_1 y_1 + \beta_2 y_2 + \beta_3 y_3 + \beta_4 z_1 + \beta_5 z_2 + \varepsilon. \qquad (26)$$

*w* is the dependent variable, usually non-financial. If financial, it may never be computed from the same financial statement data contained in $x_1$ to $x_4$, in order to prevent spurious correlations. $y_1$ to $y_3$ are the pairwise log-ratios in Equations (10) to (12), $z_1$ and $z_2$ are the non-financial predictor variables age and brand, the $\alpha$ parameter is the intercept term, and the $\beta$ parameters are effects of each of the pairwise log-ratios and non-financial predictors on *w*. These effects are interpreted keeping all other predictors (financial and non-financial) constant. The $\varepsilon$ term is the residual standing for the non-explained part of *w*.

The following six statistical hypotheses are tested corresponding to the $\beta$ parameters in the regression equation (26):

$H_0$: $\beta_1=\beta_2=\beta_3=\beta_4=\beta_5=0$ (none of the variables affects *w*),
$H_0$: $\beta_1=0$ (turnover does not affect *w*),
$H_0$: $\beta_2=0$ (margin does not affect *w*),
$H_0$: $\beta_3=0$ (leverage does not affect *w*),
$H_0$: $\beta_4=0$ (firm age does not affect *w*),
$H_0$: $\beta_5=0$ (having or not a brand does not affect *w*).

The p-value associated to each statistical test indicates the risk involved in rejecting the hypothesis. If this is low (e.g., lower than 0.05), the hypothesis can be rejected, leading to the conclusion that at least one of the predictors (global test) or the particular predictor at hand (individual test) does affect *w*, keeping all other predictors constant, in other words, that its effect is statistically significant. For instance, a positive significant $\beta_1$ coefficient would indicate that firms with a higher turnover have a higher *w*, keeping margin, leverage, age, and brand status constant. A negative significant $\beta_4$ coefficient would indicate that older firms have a lower *w*, keeping turnover, margin, leverage, and brand status constant. A negative significant $\beta_5$ coefficient would indicate that firms with a brand have a lower *w*, keeping turnover, margin, leverage, and age constant.

Since there is only one equation, there is only one set of *residual plots* to check the model's assumptions and only one $R^2$ value indicates the percentage of variance in *w* jointly explained by the financial and non-financial predictors.

The situation can be extended to a non-numeric *w* dependent variable, but this cannot be done with a linear regression model. A *generalized linear model* must be used instead (Coenders et al., 2017; Coenders and Greenacre, 2023). A useful particular case is the logistic regression model for a binary *w* variable (for instance, an indicator of default or bankruptcy coded as 1 when there is default or bankruptcy and as 0 otherwise, see Keivani et al., 2026; Magrini, 2025). Generalized linear models are outside the scope of this introduction.

To the best of our knowledge, the only published works using compositional financial ratios as explanatory variables are those by Keivani et al. (2026) and Magrini (2025). This constitutes a rich potential avenue of further research.

## 9. Conclusions

### 9.1 Summary

Compositional Data (CoDa) can be used to advantage to distil the relative magnitude of accounting figures, which lies at the ground of researchers' and professionals' interests when performing financial statement analysis. CoDa solve the main drawbacks of standard financial ratios in statistical analysis at industry level, such as the extreme skewness and outliers shown in Fig. 2, and dependence of the results on the arbitrary permutation of numerator and denominator, as shown in appendix 4.

These permutations are not uncommon in practice (Coenders et al., 2023a; Linares-Mustarós et al., 2022). Indebtedness is defined as $x_3/x_4$ and, as suggested in Equation (13) , is linked to leverage. Its inverse, called solvency, is defined as $x_4/x_3$. The concept is the same in a reversed scale. The higher the indebtedness, the higher the leverage, and the higher the solvency, the lower the leverage. In our example, regressing the standard solvency ratio thus defined on having or not a brand and firm age leads to no significant relationships ($R^2$=0.0031), while regressing the standard indebtedness ratio shows a marginally significantly lower indebtedness in firms with a brand (p-value=0.0562; $R^2$=0.0445). See appendix 4.

The common notion of average financial structure must be expressed as geometric means. Geometric means have the attractive properties in industry analysis that the mean ratio equals the ratio between means, and that the mean of the inverse ratio equals the inverse ratio between the means. Once geometric means of accounting values have been computed for the whole industry or any subdivision of it, interpretation may revert to standard financial ratios for an ease of communication with the accounting and management community. The relative statistical complexity of the log-ratios will thus be hidden from the end reader.

Classical multivariate analysis tools can be used on CoDa after transforming the data by means of centred log-ratios. In particular, the principal-component-analysis biplot visualizes all firms with respect to the ratio of any pair of accounting figures, and cluster analysis makes it possible to draw groups of firms with similar financial statement profiles. Clusters can be related to external variables in the usual way, for instance with simple mosaic plots or boxplots. Cluster analysis results can also be expressed in terms of standard financial ratios with geometric means. Submitting standard financial ratios to cluster analysis leads to some clusters being uninformative, small, or even made up only of outliers (Dao et al., 2026; Feranecová and Krigovská, 2016; Hernandez Romero and Coenders, 2025; Jofre-Campuzano and Coenders, 2022; Linares-Mustarós et al., 2018; Lukason and Laitinen, 2019; Sharma et al., 2016), In our example, submitting the standard turnover, margin, and leverage ratios to the *k*-means classification led to an uninterpretable 3-cluster solution with cluster sizes 101, 7 and 1.

The results of biplots and cluster analysis are very useful to managers. In the biplot, managers can visually compare the financial profile of their firm with that of any other firm in the industry. In the cluster analysis, managers can compare the financial profile of their firm with the average profile of a subset of similar firms, taking into account the industry heterogeneity. The comparison can be made with respect to the closest cluster at the moment of carrying out the analysis, or to the cluster the firm would aspire to belong to in the future. Sometimes these firm clusters can be identified with strategic groups, competing on the basis of margin or turnover, or with a certain choice for their capital structure.

Statistical modelling is also possible by means of $D-1$ pairwise log-ratios, which can play the role of dependent or explanatory variables in a linear regression model. Other log-ratio transformations are also candidates for this purpose, like the so-called orthonormal log-ratio coordinates -olr- a.k.a. isometric log-ratio coordinates -ilr- (Egozcue et al., 2003; Martín-Fernández, 2019; Pawlowsky-Glahn and Egozcue, 2011) used for instance by Arimany-Serrat et al. (2022; 2023); Carreras-Simó and Coenders (2021); Coenders (2025); Coenders et al. (2023a); Escaramís and Arbussà (in press); Linares-Mustarós et al. (2018; 2022); Magrini (2025); and Molas-Colomer et al. (2024). The results in Section 8 are replicated with olr in Appendix 3.

The cited current and past research in compositional financial statement analysis includes a wide range of industries, like manufacture of pharmaceutical preparations (Linares-Mustarós et al., 2018), manufacture of wearing apparel (Linares-Mustarós et al., 2018), hospitals (Creixans-Tenas et al., 2019), grocery (Carreras-Simó and Coenders, 2020), retail trade (Carreras-Simó and Coenders, 2021), wholesale trade (Keivani et al., 2026), wineries (Arimany-Serrat et al., 2022; 2023; Coenders, 2025; Linares-Mustarós et al., 2022), breweries (Arimany-Serrat and Sgorla, 2024; Coenders et al., 2023a), agriculture and farming (Molas-Colomer et al., 2026), hospitality (Farreras-Noguer et al., 2025; Mulet-Forteza et al., in press; Saus-Sala et al., 2021; 2023; 2024), beekeeping (Arimany-Serrat and Coenders, 2025), fisheries (Dao et al., 2026), food and beverage processing (Dao et al., 2026; Rondós-Casas et al., 2026), grain and fat cultivation and processing (Hernandez Romero and Coenders, 2025), and sale of automotive fuel (Jofre-Campuzano and Coenders, 2022), and has potential to expand to any other industry.

Over ten years, the accounting community has learnt that the violations of the assumptions of standard ratios can be severe, and that the improved CoDa methodology leads to results that are different enough to modify (for the better) the main conclusions of applied research. In this respect, the CoDa approach represents a vast improvement in applied accounting research using statistical methods on financial ratios. We can only encourage its widespread adoption, which should neither affect the researchers' objectives nor their taste for a particular statistical method, while interpretation can often be made in terms of either standard financial ratios or simple pairwise ratios. An introduction to compositional analysis of financial statements in French can be found in Coenders (2025), and a handbook in Spanish in Coenders and Arimany-Serrat (2025a).

CoDa is not needed when there is no substantial statistical analysis of a sample of firms. This includes case-study research, in which typically the evolution of key standard ratios is plotted over time for just one or two firms, and all qualitative research paradigms. It is not either needed to monitor the performance of one's own firm. However, CoDa is necessary to compute reliable means of the ratios at the industry level. So, comparing one's own firm to the industry average requires industry averages to be properly computed (Hernandez Romero and Coenders, 2025; Saus-Sala et al., 2021). CoDa is not needed either when using statistical methods that are robust to extreme asymmetry and outliers, which incudes but is not limited to certain non-parametric methods (Iotti et al., 2023; 2024a; 2024b; 2024c).

**9.2. Extensions**

Any positive non-overlapping accounting figures may be used to define any set of ratios, beyond the very simple DuPont-analysis case (Saus-Sala et al., 2021; 2023). For instance, current and non-current assets could have been used instead of total assets, and current and non-current liabilities could have been used instead of total liabilities bringing the number of accounting figures $D$ to 6. This would have made it possible to define ratios of asset structure (non-current over current assets), debt maturity (non-current over current liabilities), and short-term solvency (current assets over current liabilities). See Arimany-Serrat and Coenders (2025); Arimany-Serrat et al. (2023); Coenders (2025); Coenders and Arimany-Serrat (2025a); Creixans-Tenas et al. (2019); Dao et al. (2026); Farreras-Noguer et al. (2025); Hernandez Romero and Coenders (2025); Jofre-Campuzano and Coenders (2022); and Saus-Sala et al. (2024). See also Appendix 5. This notwithstanding, as stated in section 3.3, a too detailed subdivision into a large number $D$ of accounting figures may not be advisable on the grounds of large percentages of zeros, especially if the sample contains small firms. From our own experience, $D$=6 tends to be a convenient compromise. The most exhaustive study to date is that by Carreras-Simó and Coenders (2020) who use $D$=14 in a sample of large grocery distribution chains. Zeros are indeed the main threat to financial ratios (be they standard or compositional) in small firms.

To use accounting figures from other financial statements than the balance sheet and the income statement is also possible. Arimany-Serrat et al. (2022) extend the CoDa methods to the analysis of the cash-flow statement. It is also possible to include non-financial figures if they result into meaningful ratios. For instance, the number of employees appears in the average-wage ratio, the sales-per-employee ratio, the assets-per-employee ratio, and so on (Carreras-Simó and Coenders, 2020). In the hospitality sector the revenue per available room is a key management variable (Mulet-Forteza et al., in press) arguing for the inclusion of the number of rooms in the analysis. Another source of ratios is in the environmental, social and governance indicators. For instance, the ratio of energy consumption on revenues is defined as energy intensity and can be treated compositionally (Rondós-Casas et al., 2026; Todorov and Simonacci, 2020). Similar intensity ratios exist for water consumption, green-house gas emissions, waste generation, and so on.

Accounting figures can be weighted in order to equalize their impact on the results, in a similar way as standardization is used in common statistical analysis (Dao et al., 2026; Jofre-Campuzano and Coenders, 2022). Weighting can improve the biplot representation and the clustering quality. Conversely, weighting is not necessary in statistical modelling and should never be used when computing industry ratio averages. CoDa cannot be standardized in the usual manner by subtracting the mean and dividing the standard deviation.

The biplot can immediately be extended to panel data to plot trajectories of individual firms over time (Carreras-Simó and Coenders, 2020; Saus-Sala et al., 2023). In the same vein, cluster analysis on panel data can show how firms transition from one cluster to another (Arimany-Serrat and Coenders, 2025; Arimany-Serrat and Sgorla, 2024; Dao et al., 2026; Hernandez Romero and Coenders, 2025; Saus-Sala et al., 2023; 2024).

Besides the statistical methods described here, compositional financial statement analysis has used partial-least-squares structural equation modelling (Creixans-Tenas et al., 2019), vector autoregressive models (Carreras-Simó and Coenders, 2021), weighted classification (Dao et al., 2026; Jofre-Campuzano and Coenders, 2022), fuzzy classification (Molas-Colomer et al., 2024; 2026), logistic regression models (Keivani et al., 2026; Magrini, 2025), random forests (Keivani et al., 2026; Magrini, 2025), k-nearest-neighbour classifier (Keivani et al., 2026), and panel regression models (Arimany-Serrat et al., 2023; Escaramís and Arbussà, in press), and has potential to expand to any other statistical or econometrical method or model used in accounting and finance. CoDaPack does not support many of these methods but after zero replacement and log-ratio computation, the data can be exported and imported back into the researcher's favourite software. After log-ratio transformation, any statistical method can be applied in a standard manner and then interpreted from a compositional perspective. One of the strong points of the CoDa approach is that it provides a unified approach which suits itself to any statistical analysis. Beyond CoDaPack, users of R (R Core Team, 2022) can also benefit from some of the many R libraries devoted to CoDa. We especially recommend zCompositions (Palarea-Albaladejo and Martín-Fernández, 2015), compositions (van den Boogaart and Tolosana-Delgado, 2013), robCompositions (Filzmoser et al., 2018), easyCODA (Greenacre, 2018), and coda4microbiome (Calle et al., 2023).

The compositional methodology has also potential for any business research project using statistical models and including financial ratios among the set of study variables. Carreras-Simó and Coenders (2021) relate asset and capital structures, Escaramís and Arbussà (2025) compare capital structures of family and non-family firms, Keivani et al. (in press) and Magrini (2025) predict bankruptcy, Creixans-Tenas et al. (2019) study the impact of social responsibility on profitability and solvency, Mulet-Forteza et al. (in press) the impact of expansion strategies, Farreras-Noguer et al. (2025) the impact of subsidies, and Arimany-Serrat et al. (2023) the impact of Covid-19. As we have shown, compositional financial ratios can also be used as explanatory variables to predict default, firm survival or any non-financial variable.

### 9.3. Current methodological debates and challenges for further research

A current methodological debate in financial-ratio application is the choice of log-ratio transformation. In distance-based methods such as cluster analysis, olr and clr yield identical results, so that there is not much to choose between them, while there is a consensus that pairwise log-ratios are inappropriate, with some exceptions (Greenacre, 2018; 2019). olr are a powerful alternative to pairwise log-ratios in statistical modelling. In most cases, rather than being a technical matter, the choice depends on the researcher's favourite way to interpret the results. See Appendices 3 and 5 for examples.

Another current debate is the role of weights. In financial applications of CoDa methods, they have so far only been assessed on cluster analysis (Dao et al., 2026; Jofre-Campuzano and Coenders, 2022) with substantial improvements on the interpretability of the results. They have potential for other distance-based methods such as biplots. This is a promising avenue for further research.

Some statistical methods have been used only rarely in conjunction with CoDa, or not at all. This includes penalized regression (Magrini, 2025), Bayesian computation (Coenders et al., 2026), and many machine learning methods (Tran, 2025), which are becoming popular in bankruptcy prediction, with standard ratios (Nugroho and Dewayanto, 2025), and more rarely with compositional ratios (Keivani et al., 2025; Magrini, 2025). For some of these methods the best transformation is yet to be determined, which is also a promising avenue for further research.

## Declaration of conflicts of interest

The authors declare no conflict of interest.

## Data availability

The data are available in Appendix 1.

## Acknowledgements

This document is a summary of the research line «compositional analysis of accounting ratios» from the projects financed by the Spanish Ministry of Science, Innovation and Universities MCIN/AEI/10.13039/501100011033 and ERDF-a way of making Europe «METhods for COmpositional analysis of DAta (CODAMET)» [grant number RTI2018-095518-B-C21] and «GENERAtion and transfer of compositional data analysis knowledge (CODA-GENERA)» [grant number PID2021-123833OB-I00], with Salvador Linares-Mustarós, Miquel Carreras-Simó, Maria Àngels Farreras-Noguer, Germà Coenders, Xavier Molas-Colomer, Elisabet Saus-Sala and Joan Carles Ferrer-Comalat from the University of Girona (Spain), and Núria Arimany-Serrat from the University of Vic-Central University of Catalonia (Spain). Details can be found in https://www.researchgate.net/lab/Lab-on-financial-statement-analysis-as-compositional-data-Germa-Coenders-2.

This research was also supported by the Spanish Ministry of Health [grant number CIBERCB06/02/1002]; the Department of Research and Universities of Generalitat de Catalunya [grant numbers 2021SGR01197 and 2021SGR00403]; and the Department of Research and Universities, AGAUR and the Department of Climate Action, Food and Rural Agenda of Generalitat de Catalunya [grant number 2023-CLIMA-00037].

# Appendix 1: Dataset

| Firm | $x_1$ | $x_2$ | $x_3$ | $x_4$ | Brand | Age |
|---|---|---|---|---|---|---|
| 1 | 10386 | 12987 | 34048 | 41456 | 1 | 22 |
| 2 | 8004 | 8416 | 24104 | 30085 | 1 | 28 |
| 3 | 16692 | 11755 | 7440 | 101358 | 0 | 26 |
| 4 | 11510 | 10707 | 16828 | 28410 | 1 | 10 |
| 5 | 16742 | 16769 | 8579 | 22967 | 0 | 18 |
| 6 | 34840 | 32015 | 44345 | 72394 | 1 | 82 |
| 7 | 31070 | 26010 | 6234 | 35945 | 1 | 66 |
| 8 | 21140 | 17844 | 12138 | 29176 | 1 | 33 |
| 9 | 96411 | 72656 | 73509 | 324641 | 1 | 31 |
| 10 | 13421 | 11525 | 3694 | 25257 | 1 | 35 |
| 11 | 60202 | 57587 | 16335 | 71165 | 0 | 2 |
| 12 | 72582 | 65071 | 5470 | 83535 | 0 | 3 |
| 13 | 23261 | 18073 | 23986 | 56463 | 1 | 28 |
| 14 | 13941 | 13786 | 6892 | 8520 | 0 | 22 |
| 15 | 11837 | 9294 | 4655 | 25422 | 1 | 21 |
| 16 | 12452 | 9589 | 4957 | 92492 | 1 | 22 |
| 17 | 32956 | 30230 | 26594 | 66630 | 1 | 15 |
| 18 | 18307 | 13587 | 3928 | 39874 | 1 | 42 |
| 19 | 16856 | 13424 | 5149 | 55904 | 1 | 115 |
| 20 | 25625 | 23361 | 10730 | 20054 | 1 | 15 |
| 21 | 12616 | 12241 | 2835 | 6537 | 0 | 13 |
| 22 | 16968 | 14287 | 13806 | 37401 | 1 | 6 |
| 23 | 18149 | 11913 | 19651 | 35564 | 1 | 21 |
| 24 | 48717 | 32328 | 26678 | 139254 | 1 | 20 |
| 25 | 24000 | 23379 | 7775 | 13899 | 1 | 13 |
| 26 | 33709 | 32826 | 15334 | 20037 | 1 | 19 |
| 27 | 38566 | 38720 | 25590 | 28311 | 0 | 16 |
| 28 | 12260 | 12361 | 52911 | 64438 | 1 | 23 |
| 29 | 13597 | 12508 | 19147 | 35585 | 1 | 44 |
| 30 | 34434 | 30497 | 58473 | 78393 | 1 | 13 |
| 31 | 40704 | 39748 | 23538 | 32136 | 0 | 21 |
| 32 | 29981 | 28923 | 11511 | 16614 | 0 | 15 |
| 33 | 21271 | 16977 | 6748 | 64526 | 1 | 42 |
| 34 | 14932 | 11931 | 13355 | 37558 | 1 | 20 |
| 35 | 24184 | 23942 | 24460 | 32562 | 1 | 23 |
| 36 | 19633 | 18906 | 29641 | 42961 | 1 | 30 |
| 37 | 11877 | 11765 | 16962 | 18394 | 0 | 12 |
| 38 | 11288 | 11165 | 4507 | 11172 | 1 | 75 |
| 39 | 19528 | 16678 | 13119 | 42340 | 1 | 31 |
| 40 | 51184 | 45551 | 20999 | 57881 | 1 | 31 |
| 41 | 10930 | 12668 | 8845 | 25496 | 1 | 42 |
| 42 | 18795 | 18279 | 17940 | 24707 | 1 | 63 |
| 43 | 10957 | 10920 | 4594 | 10443 | 1 | 37 |
| 44 | 17586 | 16851 | 19432 | 49583 | 1 | 33 |
| 45 | 44509 | 35925 | 49157 | 97699 | 1 | 43 |
| 46 | 19018 | 17169 | 32786 | 58143 | 1 | 86 |
| 47 | 24987 | 24663 | 42050 | 50722 | 1 | 71 |
| 48 | 9598 | 8302 | 5614 | 20458 | 1 | 27 |
| 49 | 28533 | 15219 | 19654 | 122212 | 1 | 52 |
| 50 | 23628 | 23571 | 9355 | 17945 | 1 | 6 |
| 51 | 16426 | 15889 | 13243 | 64989 | 1 | 35 |
| 52 | 19549 | 19429 | 5993 | 8697 | 0 | 26 |
| 53 | 56892 | 60958 | 31041 | 59388 | 1 | 62 |
| 54 | 13024 | 12203 | 5208 | 58275 | 1 | 41 |
| 55 | 11031 | 9801 | 6031 | 8764 | 1 | 9 |
| 56 | 32692 | 32439 | 13100 | 19502 | 1 | 75 |
| 57 | 38207 | 38432 | 41987 | 58921 | 0 | 24 |
| 58 | 195083 | 198107 | 201348 | 365287 | 1 | 90 |
| 59 | 24595 | 15318 | 31384 | 195823 | 0 | 20 |
| 60 | 14124 | 14380 | 9363 | 9471 | 1 | 14 |
| 61 | 99027 | 76807 | 59139 | 183307 | 1 | 34 |
| 62 | 14321 | 14300 | 5464 | 20418 | 1 | 53 |
| 63 | 16032 | 15943 | 8013 | 16862 | 1 | 72 |
| 64 | 24123 | 23898 | 23933 | 26902 | 1 | 61 |
| 65 | 12463 | 12157 | 17251 | 19627 | 0 | 16 |
| 66 | 69561 | 54111 | 68227 | 256252 | 1 | 24 |
| 67 | 8775 | 8603 | 17703 | 23320 | 1 | 59 |
| 68 | 259734 | 240445 | 104154 | 286414 | 0 | 39 |
| 69 | 123735 | 122071 | 43130 | 116517 | 1 | 9 |
| 70 | 537137 | 526427 | 508236 | 748117 | 1 | 88 |
| 71 | 169304 | 157689 | 114442 | 329924 | 1 | 46 |
| 72 | 31456 | 31094 | 23582 | 35752 | 1 | 19 |
| 73 | 30027 | 31633 | 9857 | 66558 | 1 | 76 |
| 74 | 10786 | 10202 | 7126 | 40587 | 1 | 13 |
| 75 | 38305 | 36653 | 44082 | 53633 | 1 | 14 |
| 76 | 673107 | 617452 | 594441 | 865845 | 1 | 45 |
| 77 | 10562 | 10111 | 5004 | 27996 | 0 | 5 |
| 78 | 43787 | 38357 | 22471 | 96302 | 1 | 39 |
| 79 | 26040 | 23274 | 10976 | 23926 | 1 | 40 |
| 80 | 57164 | 51805 | 41207 | 72790 | 1 | 31 |
| 81 | 24048 | 21513 | 15137 | 45304 | 1 | 51 |
| 82 | 26784 | 25520 | 12179 | 13447 | 0 | 4 |
| 83 | 56259 | 54911 | 37037 | 55362 | 1 | 26 |
| 84 | 16940 | 17179 | 16747 | 18456 | 0 | 43 |
| 85 | 29747 | 24915 | 42917 | 77024 | 1 | 1 |
| 86 | 14779 | 11694 | 21851 | 65560 | 1 | 35 |
| 87 | 29386 | 28777 | 33508 | 44656 | 1 | 37 |
| 88 | 268730 | 257330 | 104579 | 391390 | 1 | 32 |
| 89 | 65237 | 59441 | 68706 | 97750 | 0 | 33 |
| 90 | 11314 | 7430 | 17442 | 45345 | 1 | 28 |
| 91 | 43979 | 41265 | 30812 | 86256 | 1 | 14 |
| 92 | 14970 | 14174 | 21473 | 27310 | 1 | 24 |
| 93 | 135233 | 120931 | 48661 | 221337 | 1 | 43 |
| 94 | 11751 | 16125 | 15105 | 16206 | 1 | 31 |
| 95 | 34899 | 21295 | 13763 | 100585 | 1 | 21 |
| 96 | 21327 | 21394 | 9672 | 15772 | 1 | 49 |
| 97 | 11053 | 10993 | 5787 | 9792 | 0 | 37 |
| 98 | 11584 | 9981 | 3918 | 22686 | 1 | 83 |
| 99 | 12980 | 12574 | 13922 | 17771 | 0 | 57 |
| 100 | 16625 | 16603 | 17980 | 31256 | 0 | 40 |
| 101 | 53989 | 54545 | 61072 | 81757 | 1 | 44 |
| 102 | 62461 | 60327 | 21805 | 29618 | 0 | 34 |
| 103 | 10399 | 10051 | 11465 | 19488 | 1 | 10 |
| 104 | 32673 | 29301 | 7908 | 40683 | 1 | 65 |
| 105 | 86395 | 85927 | 45882 | 55318 | 1 | 18 |
| 106 | 11862 | 10339 | 2778 | 38565 | 1 | 30 |
| 107 | 13364 | 11772 | 15292 | 46389 | 1 | 25 |
| 108 | 30172 | 29610 | 15320 | 26823 | 1 | 20 |
| 109 | 9316 | 8720 | 16050 | 21057 | 0 | 20 |

## Appendix 2: Selected CoDaPack menus

**Opening or importing data:** The *File* menu handles opening and saving data files, including importing and exporting them to a variety of formats (.xls, .csv, .txt, and .RData). File names may contain only letters in the English alphabet, numbers and underscores "_".

For instance, to import a .xls Excel file, select the *File➢Import➢Import XLS Data* menu. The Excel file must only contain one sheet with the variable names in the first row and the data from the second row onwards. Data may be text or numbers, not formulas. Variable names may contain only letters in the English alphabet, numbers, periods ".", and underscores "_", and may not include spaces. Zeros in accounting data must be entered as such; missing values in non-accounting data as "NA".

When importing the data, CoDaPack will assign variable types to the data. Numeric columns appear in white and categoric columns in yellow. The user may change that status if the variables actually contain numbers (*Data➢Manipulate➢Categoric to Numeric* or *Data➢Manipulate➢Numeric to Categoric*).

After importing the data it is advisable to store them in CoDaPack's native .cdp format (*File➢Save as* menu). To open .cdp files go to the *File➢Open Workspace* menu.

**Zero imputation:** If there are zeros in the accounting data, they have to be imputed first. The *Irregular Data➢Zero Patterns* menu, computes percentages of zeros per part and overall, and percentages of zero co-occurrence, after introducing the parts $x_1,\ldots,x_4$ into the *Selected* box with the *Show percentages* and *Plot Pattern* options.

The *Irregular Data➢Set Detection Limit* menu makes it possible to set the detection limit as the minimum value of each column after introducing the parts $x_1,\ldots,x_4$ into the *Selected* box with the *Column minimum* option. It is also possible to select any detection limit chosen by the user, for all parts or for each part separately (*Detection limit* option).

The *Irregular Data➢Log-Ratio EM Zero Replacement* menu is a convenient zero imputation method (Palarea-Albaladejo and Martín-Fernández, 2008), after introducing the parts $x_1,\ldots,x_4$ into the *Selected* box with the default options. 4 new variables free of zeros are created at the end of the data file. The *File➢Save as* menu will store the enlarged file. The procedure requires that one part has complete data for all firms and that each firm has non-zero values for at least two parts. This is easy to attain, since revenues will be all positive after removal of essential zeros.

**Fig. 2:** To plot standard financial ratios, the best alternative is to compute them first with Excel, copy and paste them as data to remove the formulas within Excel, and include them in the imported file into CoDaPack. The *Graphs➢Boxplot* menu produces the boxplots themselves after introducing the standard ratios into the *Selected* box with no options.

**Fig. 3:** The *Data➢Transformation➢ALR* menu stores the pairwise log-ratios as additional variables at the end of the data file, after introducing the two parts in the *Selected* box, the numerator part first, the denominator part last, and with the *Raw-ALR* option. The *File➢Save as* menu will store the enlarged file.

For instance, when introducing $x_1$ and $x_4$ to compute $y_1$, CoDaPack names the transformed variable alr.x1_x4. By double clicking on the variable name the user may edit it, taking into account that edited names may contain only letters in the English alphabet, numbers, periods ".", and underscores "_" without spaces. After editing each name, the return key much be pressed.

Once transformed, pairwise log-ratios can be treated with standard statistical methods. Thus, for the purpose of descriptive statistical analysis the menu *Statistics➢Classical Statistics Summary* has to be used instead of the menu *Statistics➢Compositional Statistics Summary.*

The *Graphs➢Boxplot* menu produces the boxplots themselves after introducing the pairwise log-ratios into the *Selected* box with no options.

**Fig. 4:** The *Data➢Transformation➢CLR* menu stores the centred log-ratios as additional variables at the end of the data file, after introducing all parts $x_1,\ldots, x_4$ into the *Selected* box and with the *Raw-CLR* option. The *File➢Save as* menu will store the enlarged file.

CoDaPack names the transformed variables clr.x1 to clr.x4. By double clicking on the variable names the user may edit them, taking into account that edited names may contain only letters in the English alphabet, numbers, periods ".", and underscores "_" without spaces. After editing each name, the return key must be pressed.

The *Graphs➢Boxplot* menu produces the boxplots themselves after introducing the clr variables into the *Selected* box with no options.

**Tables 2 and 3:** The *Statistics➢Compositional Statistics Summary* menu computes the compositional centre as geometric means, after introducing the parts $x_1,\ldots, x_4$ into the *Selected* box with only the *Center* option selected.

Clusters or any other subdivision within the industry can optionally be defined by selecting a categoric variable in the *Group by* box. The grouping variable must be stored as categoric (marked as yellow in the data table), or else must be previously transformed with the menu *Data➢Manipulate➢Numeric to Categoric*.

**Figs. 5 and 6:** The *Graphs ➢ CLR-biplot* menu depicts the covariance biplot. The menu computes centred log-ratios internally so that the original accounting figures $x_1,\ldots, x_4$ must be entered in the *Selected* box. Points can be coloured according to a categoric variable defining clusters or any other subdivision within the industry (*Group by* box). The grouping variable must be stored as categoric (marked as yellow in the data table), or else must be previously transformed with the menu *Data➢Manipulate➢Numeric to Categoric*.

Once the biplot is drawn, the *Data➢Show observation names* option can be used to identify individual firm points by row numbers in the data file. If the user wants points to be labelled by a variable in the data file rather than by row, he or she must first select the *Data➢Add observation names* option.

CoDaPack does not draw the pairwise log-ratios $y_1$ to $y_3$. To prepare this article they were added afterwards with a graph editing software.

**Table 5:** The *Statistics➢Multivariate Analysis➢Cluster➢K-means* menu performs *k*-means clustering and allows the user to select the desired *Number of clusters* by entering it twice in the *Minimum* and *Maximum* boxes. The menu computes appropriate log-ratios internally so that the original $x_1,\ldots, x_4$ accounting figures must be entered in the *Selected* box. The procedure displays the compositional centres by cluster (i.e., the cluster geometric means) and a new categoric variable named *Cluster* containing cluster membership is stored at the end of the data file. The *File➢Save as* menu will store the enlarged file.

An alternative possibility is to let the procedure decide the optimal number of clusters between *Minimum*=2 and a *Maximum* number of clusters decided by the user. *Optimality* may be defined by the *Average Silhouette* width or the *Caliński-Harabasz index*. CoDaPack provides plots of the Average silhouette width and the Caliński-Harabasz index and stores only the best solution in the data file as *Cluster*. It also displays the values of these statistics and the cluster geometric means.

**Fig. 7:** The *Graphs➢Mosaic plot* menu draws the mosaic plot. Two variables have to be entered in the *Selected* box, the one in the horizontal axis first. The variables must be stored as categoric (marked as yellow in the data table), or else must be previously transformed with the menu *Data➢Manipulate➢Numeric to Categoric*.

**Fig. 8:** The *Graphs➢Boxplot* menu produces the boxplot, after introducing the firm age into the *Selected* box with no options and selecting the cluster variable for separated boxplots in the *Group by* box.

**Fig. 9:** The *Graphs➢Boxplot* menu produces the boxplot, after introducing the same pairwise log-ratios constructed for Fig. 3 into the *Selected* box with no options and selecting the categoric variable for separated boxplots in the *Group by* box.

**Fig. 10:** The *Graphs➢Scatterplot* menu produces scatterplots by introducing two numeric variables in the *Selected* box. The variable entered first appears in the horizontal axis.

**Table 7 and Fig. 11:** Once transformed as log-ratios, compositional data become real values between minus and plus infinity. The *Statistics➢Multivariate Analysis➢Regression➢X real Y real* menu performs linear regression and draws from the previously transformed pairwise log-ratios $y_1,\ldots, y_3$, which have to be introduced one at a time in the *Response variable* box, and the numeric predictors, which have to be entered all together in the *Explanatory variables* box.

If stored as categoric, binary predictor variables have beforehand been coded as 0 and 1 (*Data➢Manipulate➢Change Category Labels* menu) and then declared as numeric in the *Data➢Manipulate➢Categoric to Numeric* menu.

## Appendix 3: Orthonormal log-ratio coordinates

The so-called *isometric log-ratio (ilr) coordinates* (Egozcue et al., 2003), recently better known under the term *orthonormal log-ratio (olr) coordinates* (Martín-Fernández, 2019), are recommended on the grounds that they are usable in virtually any statistical analysis besides being interpretable in accounting (Arimany-Serrat et al., 2022; 2023; Carreras-Simó and Coenders, 2021; Coenders, 2025; Coenders et al., 2023a; Escaramís and Arbussà, 2025; Linares-Mustarós et al., 2018; 2022; Magrini, 2025; Molas-Colomer et al., 2024). They can thus be used in any case in which pairwise log-ratios can be used. Their only drawback is their greater conceptual complexity. In this appendix we show their use as dependent variables in statistical modelling, following Section 8.1.

Interpretable olr coordinates can be easily formed from a *sequential binary partition* (SBP) of parts (Egozcue and Pawlowsky-Glahn, 2005; Pawlowsky-Glahn and Egozcue, 2011). To create the first olr coordinate, the complete composition $\mathbf{x}=(x_1,x_2,\ldots,x_D)$ is split into two groups of parts: one for the numerator and the other for the denominator of the log-ratio. In the following step, one of the two groups is further split into two subgroups to create the second olr coordinate. In any step of the SBP, when the $y_j$ olr coordinate is created, a group containing $r+s$ parts resulting from one of the previous partitions is split into two: $r$ parts $(x_{n1},\ldots, x_{nr})$ are placed in the numerator, and $s$ parts $(x_{d1},\ldots,x_{ds})$ in the denominator. The olr coordinate is a scaled log-ratio of the geometric means of each group of parts:

$$y_j = \sqrt{\frac{r\,s}{r+s}}\,\log\frac{\sqrt[r]{x_{n1}\ldots x_{nr}}}{\sqrt[s]{x_{d1}\ldots x_{ds}}}\,. \tag{27}$$

The greater the coordinate, the greater the importance of the parts (accounting figures) in the numerator as compared to those in the denominator. $\sqrt{\frac{r\,s}{r+s}}$ is only a scaling constant used to take the number of parts involved into account without changing the interpretation of the coordinate. It must be noted that $D$ parts always result in only $D-1$ coordinates.

It is advisable to choose a SBP which can be interpreted in the light of the accounting and management concepts of interest, which lends itself to building olr analogues to known standard ratios such as those involved in DuPont analysis. The SBP is commonly expressed with a sign matrix, in which positive signs indicate parts in the numerator and negative signs parts in the denominator. Blank cells indicate parts which are neither in the numerator nor in the denominator. Note that all parts are involved in the first partition leading to $y_1$, and only subsets of parts appear thereafter.

| | $y_1$ | $y_2$ | $y_3$ |
|---|---|---|---|
| $x_1$: revenues | + | + | |
| $x_2$: costs | + | − | |
| $x_3$: liabilities | − | | + |
| $x_4$: assets | − | | − |

$$\tag{28}$$

At the start of the SBP the $y_1$ coordinate balances revenues and costs with assets and liabilities. More precisely, $y_1$ can be formulated in several ways:

$$y_1 = \sqrt{\frac{4}{4}} \log \frac{\sqrt[2]{x_1 x_2}}{\sqrt[2]{x_3 x_4}} = \log\left( \frac{\sqrt[2]{x_1}}{\sqrt[2]{x_4}} \frac{\sqrt[2]{x_2}}{\sqrt[2]{x_3}} \right) =$$
$$\log\left( \sqrt[2]{\frac{x_1}{x_4}} \right) + \log\left( \sqrt[2]{\frac{x_2}{x_3}} \right) = \frac{1}{2} \log \frac{x_1}{x_4} + \frac{1}{2} \log \frac{x_2}{x_3}. \quad (29)$$

The higher the $y_1$ coordinate, the higher the turnover $x_1/x_4$. A higher $y_1$ figure also shows shorter cost payment cycles ($x_2/x_3$). Altogether it makes sense as a generalized turnover indicator which takes costs and liabilities into account and not only revenues and assets. Note that in a log scale the geometric mean is related to the sum normally used in standard ratios: $2\log\left(\sqrt[2]{x_1 x_2}\right) = \log(x_1) + \log(x_2)$. Note also the way in which the scaling constant is computed, as there are $r$=2 parts in the numerator and $s$=2 parts in the denominator:

$$\sqrt{\frac{2 \times 2}{2+2}} = \sqrt{\frac{4}{4}}. \quad (30)$$

The second partition compares revenues and costs (the two parts in the numerator of the previous partition) and the resulting $y_2$ coordinate is just a scaled version of the margin indicator constructed as a pairwise log-ratio in Equation (11):

$$y_2 = \sqrt{\frac{1}{2}} \log\left( \frac{x_1}{x_2} \right). \quad (31)$$

Note the way in which the scaling constant is computed, as there are $r$=1 parts in the numerator and $s$=1 parts in the denominator:

$$\sqrt{\frac{1 \times 1}{1+1}} = \sqrt{\frac{1}{2}}. \quad (32)$$

The third partition compares assets and liabilities (the two parts in the denominator of the first partition) and the resulting $y_3$ coordinate is just a scaled version of the leverage indicator constructed as a pairwise log-ratio in Equation (12):

$$y_3 = \sqrt{\frac{1}{2}} \log\left( \frac{x_3}{x_4} \right). \quad (33)$$

The boxplots in Fig. 12 relate the olr coordinates to the brand variable. Note that those for $y_2$ and $y_3$ are identical to Fig. 9 except for the scale of the vertical axis. It must be noted that the partition always leads to at least one coordinate which is just a scaled pairwise log-ratio and can be interpreted as such, in this case two of them, $y_2$ and $y_3$ (Hron et al., 2021). It goes without saying that scatterplots with firm age as in Fig. 10 could and should also be drawn.

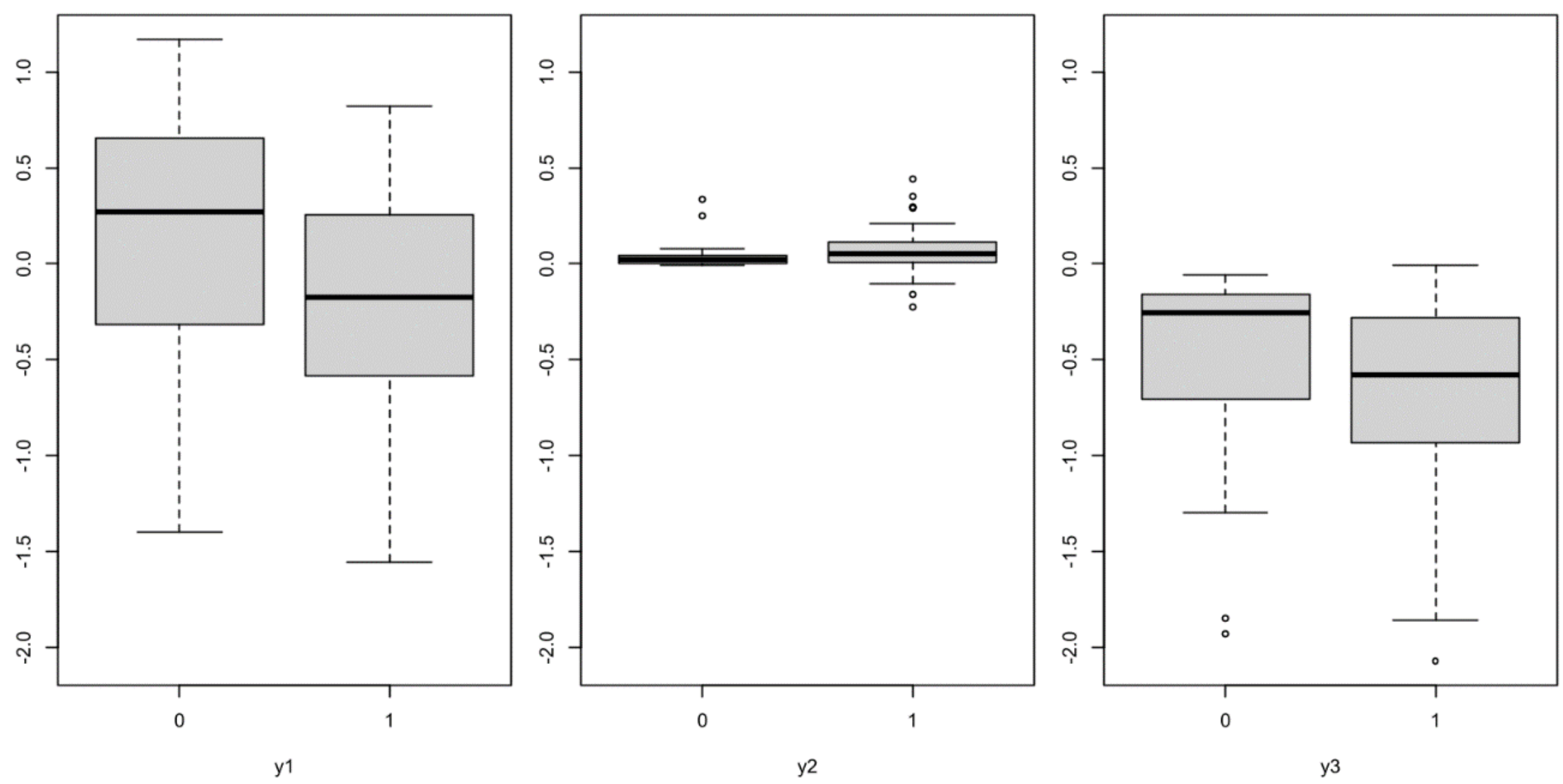

**Figure 12.** Boxplots of olr coordinates in Equations (29), (31), and (33) for wineries with (1) and without (0) their own brand

Table 8 replicates the analysis in Equation (25) and Table 7 with the coordinates in Equations (29), (31) and (33). Results of $y_2$ and $y_3$ are identical to those in Table 7 up to a scaling of the $\beta$ estimates: age and having or not a brand do not significant affect margin and leverage. Results of $y_1$ are broadly similar to Table 7: only turnover is significantly different depending on whether firms have or fail to have an own brand (p-value=0.0117). The negative sign of the coefficient (–0.3357) indicates that firms with a brand (labelled as 1) have a lower turnover, keeping firm age constant.

| | Age ($z_1$) | | Brand ($z_2$) | | | Global |
|---|---|---|---|---|---|---|
| | $\beta$ estimate | p-value | $\beta$ estimate | p-value | $R^2$ | p-value |
| $y_1$ (turnover) | 0.0010 | 0.6699 | –0.3357 | 0.0117 | 0.0592 | 0.0395 |
| $y_2$ (margin) | –0.0004 | 0.4004 | 0.0316 | 0.1762 | 0.0194 | 0.3541 |
| $y_3$ (leverage) | –0.0013 | 0.5469 | –0.1322 | 0.2664 | 0.0198 | 0.3470 |

**Table 8.** Regression estimates for the olr coordinates predicted by firm age and the variable indicating wineries with their own brand

As regards the implementation with CoDaPack:

**Fig. 12:** The *Data➢Transformation➢Raw-OLR* menu stores the orthonormal log-ratio (olr) coordinates as additional variables, after introducing the parts $x_1$,…, $x_4$ into the *Selected* box. Under *Options➢Define Manually*, one must enter the SBP. By clicking a cell in the sign matrix, the sign changes from negative (denominator) to positive (numerator) or vice-versa. One moves between columns (i.e., between coordinates) with the *Previous* and *Next* buttons. This makes it possible to draw boxplots of olr coordinates.

**Table 8:** The *Statistics➢Multivariate Analysis➢Regression➢X real, Y composition* menu performs linear regression with olr coordinates and draws from the original (i.e., not yet real-valued) accounting figures $x_1$,…, $x_4$ (entered all together in the *Response composition* box), and the numeric predictors (entered all together in the *Explanatory variables* box). The SBP has to be selected in the *Manual* button. By clicking a cell in the sign matrix, the sign changes from negative (denominator) to positive (numerator) or vice-versa. One moves between columns with the *Previous* and *Next* buttons.

## Appendix 4: Results for standard indebtedness and leverage ratios

This appendix shows how standard financial ratios can be affected by outliers, asymmetry and permutation of numerator and denominator in a regression model (Linares-Mustarós et al., 2022; Coenders et al., 2023a). As dependent variables (section 8.1), we consider five standard and compositional ratios connected to comparing assets and liabilities, that is to say, to the broad concept of indebtedness and leverage:

- Standard leverage ratio in Equation (5): $x_4/(x_4–x_3)$.
- Standard solvency ratio: $x_4/x_3$.
- Standard indebtedness ratio: $x_3/x_4$.
- Compositional leverage ratio as the pairwise log-ratio in Equation (12): $y_3=\log(x_3/x_4)$.
- The inverse of the compositional leverage ratio in Equation (12): $–y_3=\log(x_4/x_3)$.

The boxplots in Fig. 13 show strong asymmetry and many outliers in the standard ratios for leverage and solvency, and a particularly extreme one in leverage. Although indebtedness is just solvency after numerator and denominator permutation, the aspect of the boxplots and their outliers are completely different. Conversely, the compositional ratios get just the same pattern upside down after permutation.

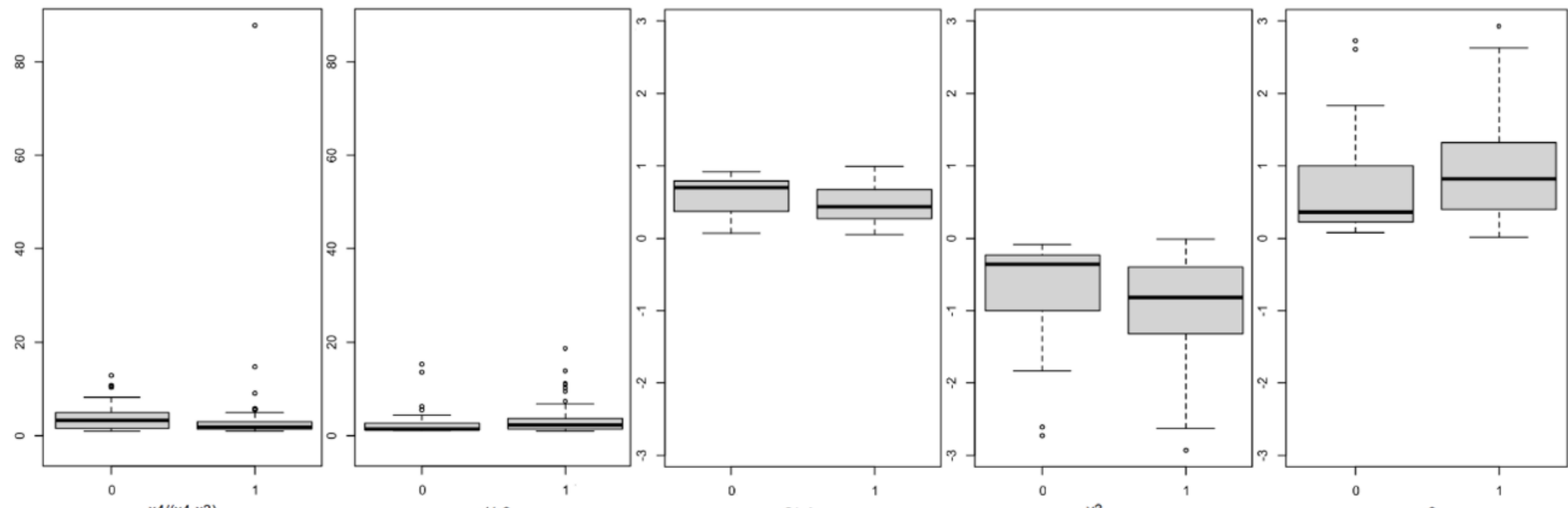

**Figure 13.** Boxplots for selected standard financial ratios and pairwise log-ratios for wineries with (1) and without (0) their own brand

| | Age ($z_1$) | | Brand ($z_2$) | | | Global |
|---|---|---|---|---|---|---|
| | $\beta$ estimate | p-value | $\beta$ estimate | p-value | $R^2$ | p-value |
| $x_4/(x_4–x_3)$ | –0.0440 | 0.2519 | –0.1501 | 0.9417 | 0.0139 | 0.4776 |
| $x_4/x_3$ | 0.0067 | 0.6427 | 0.1581 | 0.8390 | 0.0031 | 0.8464 |
| $x_3/x_4$ | –0.0006 | 0.5961 | –0.1170 | 0.0562 | 0.0445 | 0.0898 |
| $y_3=\log(x_3/x_4)$ | –0.0019 | 0.5469 | –0.1869 | 0.2664 | 0.0198 | 0.1096 |
| $–y_3=\log(x_4/x_3)$ | 0.0019 | 0.5469 | 0.1869 | 0.2664 | 0.0198 | 0.1096 |

**Table 9.** Regression estimates for selected standard financial ratios and pairwise log-ratios predicted by firm age and the variable indicating wineries with their own brand

Table 9 shows the regression results with the age and brand variables. We want to note that the results for the pairwise log-ratio ($y_3$) are equivalent after permutation ($–y_3$) and correspond to Table 7. Only the coefficient sign changes. This is not the case for the standard ratios of solvency ($x_4/x_3$) and indebtedness ($x_3/x_4$). Some results between both are markedly different ($R^2$ and the p-value for the own-brand variable, which nearly touches statistical significance when using indebtedness). The results for the standard leverage ratio $x_4/(x_4{-}x_3)$ are affected by the outlier, as shown below.

The residual plots in Fig. 14 provide examples of violations of the regression model assumptions when using standard ratios. The plots with the leverage ratio show a very extreme outlier in all plots, which is identified as firm 60, whose Cook's distance is larger than 0.5 as revealed by the residuals vs. leverage plot. The Q-Q plot for solvency shows extreme non-normality, which is not the case for the inverted indebtedness ratio. On the contrary, the sets of plots for the pairwise log-ratios are just mirror images of each other.

The estimation of regression models with standard ratios can be done in the usual manner with the *Statistics*➢*Multivariate Analysis*➢*Regression*➢*X real Y real* menu, once the standard ratios are in the data file.

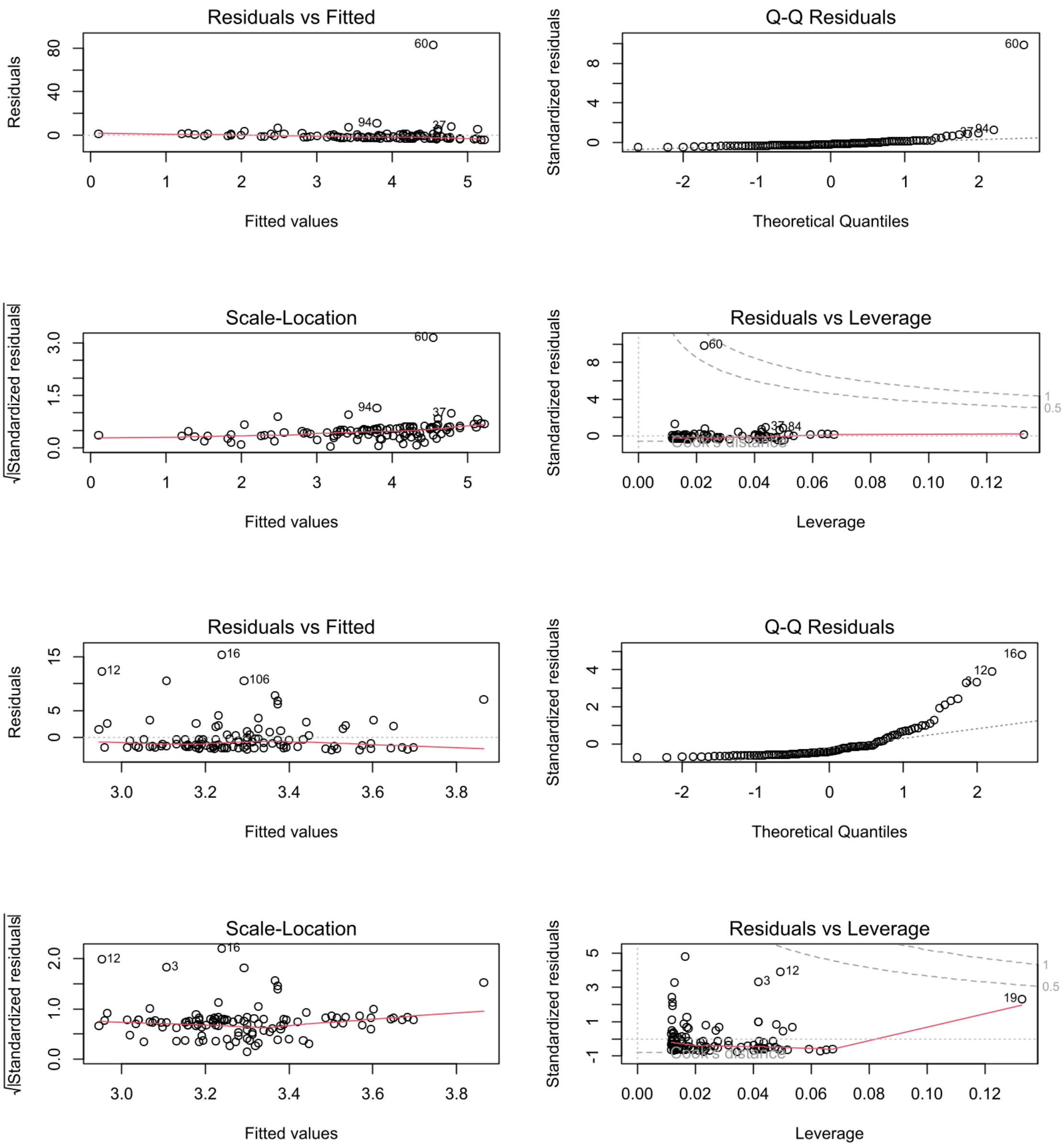

**Figure 14.** Residual plots for the equations predicting leverage ($x_4/(x_4{-}x_3)$) -top-, and solvency ($x_4/x_3$) -bottom-

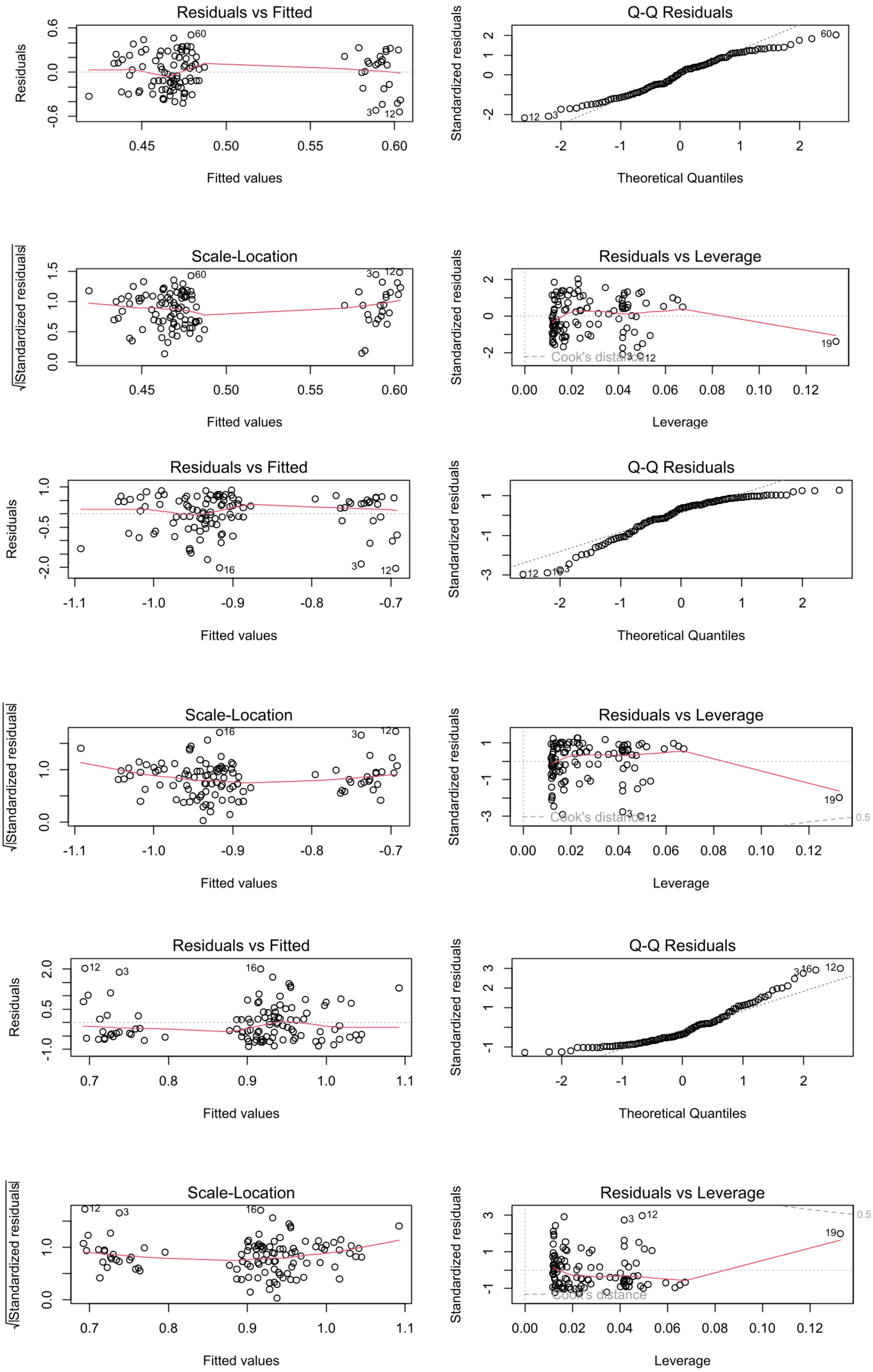

**Figure 14 continued.** Residual plots for the equations predicting indebtedness $x_3/x_4$ -top-, compositional leverage ($\log(x_3/x_4)$) -middle-, and its inverse ($\log(x_4/x_3)$) -bottom-

## Appendix 5: A proposal with *D*=6 accounting figures

In research published so far, the most common applications use the following *D*=6 positive and non-overlapping account categories as parts in the composition (Arimany-Serrat and Coenders, 2025; Arimany-Serrat et al., 2023; Coenders, 2025; Coenders and Arimany-Serrat, 2025a; Creixans-Tenas et al., 2019; Dao et al., 2026; Farreras-Noguer et al., 2026; Hernandez Romero and Coenders, 2026; Jofre-Campuzano and Coenders, 2022; Saus-Sala et al., 2024):

- $x_1$: Non-current assets,
- $x_2$: Current assets,
- $x_3$: Non-current liabilities,
- $x_4$: Current liabilities,
- $x_5$: Revenues,
- $x_6$: Costs.

These account categories are very relevant in practice because they make it possible to compute some of the most common standard ratios of turnover, margin, leverage, *long- and short-term solvency*, *asset structure*, and *debt maturity*, used in financial health and financial performance assessment:

- Turnover:

$$\text{Revenues over total assets}=x_5/(x_1+x_2). \tag{34}$$

- Current-asset turnover:

$$\text{Revenues over current assets}=x_5/x_2. \tag{35}$$

- Margin:

$$\text{Profit over revenues}=(x_5-x_6)/x_5. \tag{36}$$

- Leverage:

$$\text{Assets over equity}=(x_1+x_2)/(x_1+x_2-x_3-x_4). \tag{37}$$

- Return on assets (ROA):

$$\text{Profit over assets}=(x_5-x_6)/(x_1+x_2). \tag{38}$$

- Return on equity (ROE):

$$\text{Profit over equity}=(x_5-x_6)/(x_1+x_2-x_3-x_4). \tag{39}$$

- Indebtedness:

$$\text{Liabilities over assets}=(x_3+x_4)/(x_1+x_2). \tag{40}$$

- Indebtedness (short term):

  Current liabilities over assets=$x_4/(x_1+x_2)$. (41)

- Solvency (long-term):

  Assets over liabilities=$(x_1+x_2)/(x_3+x_4)$. (42)

- Solvency (short-term), liquidity ratio or current ratio:

  Current assets over current liabilities=$x_2/x_4$. (43)

- Asset structure or asset tangibility:

  Non-current assets over current assets=$x_1/x_2$. (44)

- Debt maturity:

  Non-current liabilities over current liabilities=$x_3/x_4$. (45)

The ratios in Equations (34) to (45) can be used to compute industry or cluster averages from the corresponding geometric means (Arimany-Serrat and Coenders, 2025; Coenders, 2025; Dao et al., 2026; Hernandez Romero and Coenders, 2026; Jofre-Campuzano and Coenders, 2022; Saus-Sala et al., 2024).

Some meaningful pairwise log-ratios are related to the standard ratios listed above, define the connected acyclic graph in Fig. 15 (top panel) and can be used in statistical modelling (Creixans-tenas et al., 2019). It must be remembered that arrows point at the numerator parts. Current asset turnover compares revenues and current assets:

$$y_1 = \log\left(\frac{x_5}{x_2}\right). \quad (46)$$

Comparing revenues and costs provides a notion of margin:

$$y_2 = \log\left(\frac{x_5}{x_6}\right). \quad (47)$$

Comparing current assets and current liabilities indicates short-term solvency:

$$y_3 = \log\left(\frac{x_2}{x_4}\right). \quad (48)$$

Comparing non-current and current assets indicates asset structure:

$$y_4 = \log\left(\frac{x_1}{x_2}\right). \quad (49)$$

Comparing non-current and current liabilities indicates debt maturity:

$$y_5 = \log\left(\frac{x_3}{x_4}\right). \quad (50)$$

As an example of redundant log-ratio choice, one could consider adding $y_6 = \log(x_3/x_1)$ to indicate to what extent non-current assets are being financed by non-current liabilities. This creates a cycle in the graph (Fig. 15, centre panel). There are two ways of joining $x_4$ and $x_1$: through $x_2$ and through $x_3$ (remember that it is not necessary to follow the arrow directions). Besides, $x_1$, $x_2$, $x_4$, $x_3$, and $x_1$ define a closed cycle. Redundancy is also shown by the fact that $y_6$ is contained in the other log-ratios. In particular, $y_6$ can be obtained as $y_5 - y_4 - y_3$:

$$\begin{aligned} & y_5 - y_4 - y_3 = \log\left(\frac{x_3}{x_4}\right) - \log\left(\frac{x_1}{x_2}\right) - \log\left(\frac{x_2}{x_4}\right) = \\ & \log(x_3) - \log(x_4) - \left(\log(x_1) - \log(x_2)\right) - \left(\log(x_2) - \log(x_4)\right) = \\ & \log(x_3) - \log(x_1) = \log\left(\frac{x_3}{x_1}\right) = y_6. \end{aligned} \quad (51)$$

The bottom panel of Fig. 15 shows an example of non-connected graph, even if the number of log-ratios is correct at $D-1=5$. There is no way of joining, for instance, $x_1$ and $x_6$. The cycle persists.

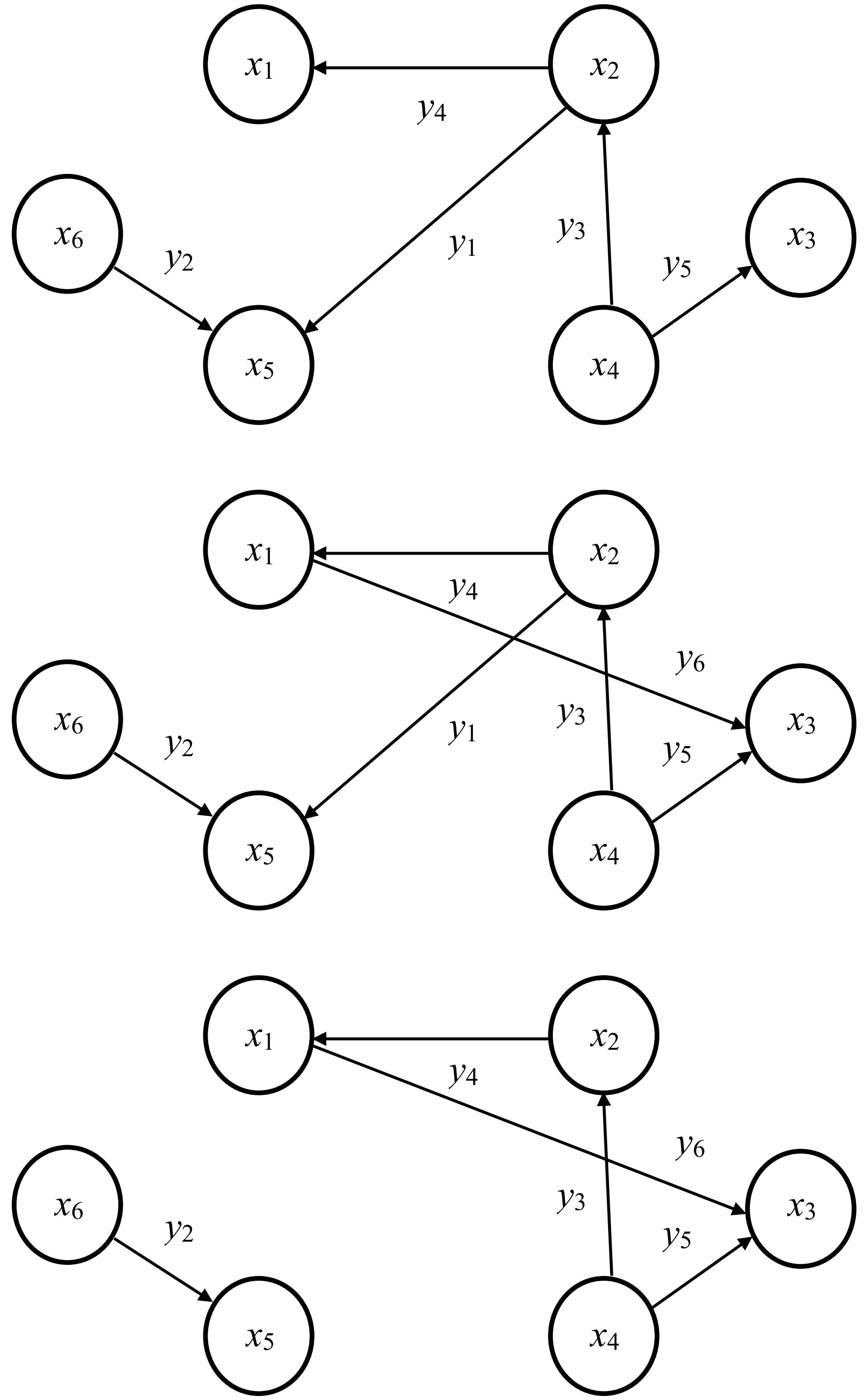

**Figure 15.** Connected acyclic graph (top), connected cyclic graph (centre), unconnected cyclic graph (bottom)

The centred log-ratios used as a basis for biplots and for clustering are (Coenders, 2025; Dao et al., 2026; Hernandez Romero and Coenders, 2026; Jofre-Campuzano and Coenders, 2022; Saus-Sala et al., 2024):

$$clr_1 = \log\left(\frac{x_1}{\sqrt[6]{x_1 x_2 x_3 x_4 x_5 x_6}}\right)$$
$$clr_2 = \log\left(\frac{x_2}{\sqrt[6]{x_1 x_2 x_3 x_4 x_5 x_6}}\right)$$
$$clr_3 = \log\left(\frac{x_3}{\sqrt[6]{x_1 x_2 x_3 x_4 x_5 x_6}}\right)$$
$$clr_4 = \log\left(\frac{x_4}{\sqrt[6]{x_1 x_2 x_3 x_4 x_5 x_6}}\right) \quad (52)$$
$$clr_5 = \log\left(\frac{x_5}{\sqrt[6]{x_1 x_2 x_3 x_4 x_5 x_6}}\right)$$
$$clr_6 = \log\left(\frac{x_6}{\sqrt[6]{x_1 x_2 x_3 x_4 x_5 x_6}}\right).$$

All possible pairwise log-ratios are contained in the centred log-ratios. Note, for instance, how $y_1$ can be obtained from $clr_5$ and $clr_2$:

$$clr_5 - clr_2 = \log\left(\frac{x_5}{\sqrt[6]{x_1 x_2 x_3 x_4 x_5 x_6}}\right) - \log\left(\frac{x_2}{\sqrt[6]{x_1 x_2 x_3 x_4 x_5 x_6}}\right) =$$
$$\log(x_5) - \log\left(\sqrt[6]{x_1 x_2 x_3 x_4 x_5 x_6}\right) - \left(\log(x_2) - \log\left(\sqrt[6]{x_1 x_2 x_3 x_4 x_5 x_6}\right)\right) = \quad (53)$$
$$\log(x_5) - \log(x_2) = \log\left(\frac{x_5}{x_2}\right) = y_1.$$

The following SBP can be used to define a set of interpretable olr coordinates that can be used in statistical modelling (Arimany-Serrat et al., 2023; Coenders, 2025):

| | $y_1$ | $y_2$ | $y_3$ | $y_4$ | $y_5$ |
|---|---|---|---|---|---|
| $x_1$: Non-current assets | − | | + | + | |
| $x_2$: Current assets | − | | + | − | |
| $x_3$: Non-current liabilities | − | | − | | + |
| $x_4$: Current liabilities | − | | − | | − |
| $x_5$: Revenues | + | + | | | |
| $x_6$: Costs | + | − | | | |

(54)

At the top of the SBP the $y_1$ coordinate balances revenues and costs with assets and liabilities. Altogether it makes sense as a turnover indicator like Equation (29). It must be remembered that positive signs indicate parts in the numerator:

$$y_1 = \sqrt{\frac{8}{6}} \log \frac{\sqrt[2]{x_5 x_6}}{\sqrt[4]{x_1 x_2 x_3 x_4}}. \quad (55)$$

The second partition compares revenues and costs, and the resulting $y_2$ coordinate is just a scaled version of the margin indicator in Equation (47):

$$y_2 = \sqrt{\frac{1}{2}} \log\left(\frac{x_5}{x_6}\right). \quad (56)$$

The third partition compares assets and liabilities, and the resulting $y_3$ coordinate is an indicator of long-term solvency:

$$y_3 = \sqrt{\frac{4}{4}} \log \frac{\sqrt[2]{x_1 x_2}}{\sqrt[2]{x_3 x_4}}. \quad (57)$$

The fourth partition compares non-current assets with current assets, and the resulting $y_4$ coordinate is just a scaled version of the asset-structure indicator in Equation (49):

$$y_4 = \sqrt{\frac{1}{2}} \log\left(\frac{x_1}{x_2}\right). \quad (58)$$

The fifth partition compares non-current liabilities with current liabilities, and the resulting $y_5$ coordinate is just a scaled version of the debt-maturity indicator in Equation (50):

$$y_5 = \sqrt{\frac{1}{2}} \log\left(\frac{x_3}{x_4}\right). \quad (59)$$

Note that pairwise log-ratios and olr coordinates not always can express equivalent financial concepts. For instance, long-term solvency in Equation (57) involves four accounting figures and cannot be expressed by means of a pairwise log-ratio. Conversely, short-term solvency in Equation (48) could have been expressed as an olr coordinate by modifying the way parts are subdivided in the SBP (Hron et al., 2021). For instance, short-term solvency is $y_5$ in the following SBP:

| | $y_1$ | $y_2$ | $y_3$ | $y_4$ | $y_5$ |
|---|---|---|---|---|---|
| $x_1$: Non-current assets | − | | + | + | |
| $x_2$: Current assets | − | | − | | + |
| $x_3$: Non-current liabilities | − | | + | − | |
| $x_4$: Current liabilities | − | | − | | − |
| $x_5$: Revenues | + | + | | | |
| $x_6$: Costs | + | − | | | |

(60)

If the researcher is interested both in the log-ratios in Equations (54) and (60), the statistical model of interest can be rerun twice.

Other sets of accounting figures can be found in Arimany-Serrat and Sgorla (2024); Carreras-Simó and Coenders (2020; 2021); Escaramís and Arbussà (2025); Linares-Mustarós et al. (2018; 2022); and Magrini (2025), which speaks for the flexibility of the CoDa approach to constructing statistical variables based on accounting-statement data.